\documentclass[journal=jpcbfk, manuscript=article]{achemso}

\usepackage{listings}
\usepackage{fancyvrb}
\usepackage[flushleft]{threeparttable}
\DefineShortVerb{\|}

\usepackage{color}
\definecolor{gray}{rgb}{0.7,0.7,0.7}
\definecolor{darkblue}{rgb}{0.0,0.0,0.6}
\definecolor{cyan}{rgb}{0.0,0.5,0.7}

\newcommand\Smaller{\fontsize{9}{9.2}\selectfont}

\lstdefinelanguage{XML}
{
  basicstyle=\Smaller\ttfamily,
  frame=<topline|bottomline>,
  morestring=[b]",
  morestring=[s]{>}{<},
  morecomment=[s]{<?}{?>},
  stringstyle=\color{black},
  identifierstyle=\color{darkblue},
  keywordstyle=\color{cyan},
  morekeywords={name, class, element, mass, class1, class2, class3, class4, type1, type2, type3,type4, c1,c2,c3,c4,c5,c6, sigma, epsilon, k, angle, length, charge}
}

\definecolor{maroon}{cmyk}{0, 0.87, 0.68, 0.32}
\definecolor{halfgray}{gray}{0.55}
\definecolor{ipython_frame}{RGB}{207, 207, 207}
\definecolor{ipython_bg}{RGB}{247, 247, 247}
\definecolor{ipython_red}{RGB}{186, 33, 33}
\definecolor{ipython_green}{RGB}{0, 128, 0}
\definecolor{ipython_cyan}{RGB}{64, 128, 128}
\definecolor{ipython_purple}{RGB}{170, 34, 255}

\lstdefinelanguage{python}{
    morekeywords={access,and,break,class,continue,def,del,elif,else,except,exec,finally,for,from,global,if,import,in,is,lambda,not,or,pass,print,raise,return,try,while,as},
    morekeywords=[2]{abs,all,any,basestring,bin,bool,bytearray,callable,chr,classmethod,cmp,compile,complex,delattr,dict,dir,divmod,enumerate,eval,execfile,file,filter,float,format,frozenset,getattr,globals,hasattr,hash,help,hex,id,input,int,isinstance,issubclass,iter,len,list,locals,long,map,max,memoryview,min,next,object,oct,open,ord,pow,property,range,raw_input,reduce,reload,repr,reversed,round,set,setattr,slice,sorted,staticmethod,str,sum,super,tuple,type,unichr,unicode,vars,xrange,zip,apply,buffer,coerce,intern},
    sensitive=true,
    morecomment=[l]\#,
    morestring=[b]',
    morestring=[b]",
    morestring=[s]{'''}{'''},
    morestring=[s]{"""}{"""},
    morestring=[s]{r'}{'},
    morestring=[s]{r"}{"},
    morestring=[s]{r'''}{'''},
    morestring=[s]{r"""}{"""},
    morestring=[s]{u'}{'},
    morestring=[s]{u"}{"},
    morestring=[s]{u'''}{'''},
    morestring=[s]{u"""}{"""},
    literate=
    {á}{{\'a}}1 {é}{{\'e}}1 {í}{{\'i}}1 {ó}{{\'o}}1 {ú}{{\'u}}1
    {Á}{{\'A}}1 {É}{{\'E}}1 {Í}{{\'I}}1 {Ó}{{\'O}}1 {Ú}{{\'U}}1
    {à}{{\`a}}1 {è}{{\`e}}1 {ì}{{\`i}}1 {ò}{{\`o}}1 {ù}{{\`u}}1
    {À}{{\`A}}1 {È}{{\'E}}1 {Ì}{{\`I}}1 {Ò}{{\`O}}1 {Ù}{{\`U}}1
    {ä}{{\"a}}1 {ë}{{\"e}}1 {ï}{{\"i}}1 {ö}{{\"o}}1 {ü}{{\"u}}1
    {Ä}{{\"A}}1 {Ë}{{\"E}}1 {Ï}{{\"I}}1 {Ö}{{\"O}}1 {Ü}{{\"U}}1
    {â}{{\^a}}1 {ê}{{\^e}}1 {î}{{\^i}}1 {ô}{{\^o}}1 {û}{{\^u}}1
    {Â}{{\^A}}1 {Ê}{{\^E}}1 {Î}{{\^I}}1 {Ô}{{\^O}}1 {Û}{{\^U}}1
    {œ}{{\oe}}1 {Œ}{{\OE}}1 {æ}{{\ae}}1 {Æ}{{\AE}}1 {ß}{{\ss}}1
    {ç}{{\c c}}1 {Ç}{{\c C}}1 {ø}{{\o}}1 {å}{{\r a}}1 {Å}{{\r A}}1
    {€}{{\EUR}}1 {£}{{\pounds}}1
    {^}{{{\color{ipython_purple}\^{}}}}1
    {=}{{{\color{ipython_purple}=}}}1
    {+}{{{\color{ipython_purple}+}}}1
    {*}{{{\color{ipython_purple}$^\ast$}}}1
    {/}{{{\color{ipython_purple}/}}}1
    {+=}{{{+=}}}1
    {-=}{{{-=}}}1
    {*=}{{{$^\ast$=}}}1
    {/=}{{{/=}}}1,
    literate=
    *{-}{{{\color{ipython_purple}-}}}1
     {?}{{{\color{ipython_purple}?}}}1,
    identifierstyle=\color{black}\ttfamily,
    commentstyle=\color{ipython_cyan}\ttfamily,
    stringstyle=\color{ipython_red}\ttfamily,
    keepspaces=true,
    showspaces=false,
    showstringspaces=false,
    rulecolor=\color{ipython_frame},
    frame=single,
    frameround={t}{t}{t}{t},
    framexleftmargin=6mm,
    numbers=left,
    numberstyle=\tiny\color{halfgray},
    backgroundcolor=\color{ipython_bg},
    basicstyle=\scriptsize,
    keywordstyle=\color{ipython_green}\ttfamily,
}

\title{Formalizing Atom-typing and the Dissemination of Force Fields with Foyer}

\author{Christoph Klein}
\affiliation{Department of Chemical and Biomolecular Engineering, Vanderbilt University, Nashville, Tennessee 37235, United States}
\alsoaffiliation{Vanderbilt Multiscale Modeling and Simulation (MuMS) Center, Vanderbilt University, Nashville, Tennessee 37235, USA}

\author{Andrew Z. Summers}
\affiliation{Department of Chemical and Biomolecular Engineering, Vanderbilt University, Nashville, Tennessee 37235, United States}
\alsoaffiliation{Vanderbilt Multiscale Modeling and Simulation (MuMS) Center, Vanderbilt University, Nashville, Tennessee 37235, USA}

\author{Matthew W. Thompson}
\affiliation{Department of Chemical and Biomolecular Engineering, Vanderbilt University, Nashville, Tennessee 37235, United States}
\alsoaffiliation{Vanderbilt Multiscale Modeling and Simulation (MuMS) Center, Vanderbilt University, Nashville, Tennessee 37235, USA}

\author{Justin Gilmer}
\affiliation{Interdisciplinary Materials Science Program, Vanderbilt University, Nashville, Tennessee 37235, United States}
\alsoaffiliation{Vanderbilt Multiscale Modeling and Simulation (MuMS) Center, Vanderbilt University, Nashville, Tennessee 37235, USA}

\author{Clare McCabe}
\affiliation{Department of Chemical and Biomolecular Engineering, Vanderbilt University, Nashville, Tennessee 37235, United States}
\alsoaffiliation{Vanderbilt Multiscale Modeling and Simulation (MuMS) Center, Vanderbilt University, Nashville, Tennessee 37235, USA}
\alsoaffiliation{Department of Chemistry, Vanderbilt University, Nashville, Tennessee 37235, United States}

\author{Peter T. Cummings}
\affiliation{Department of Chemical and Biomolecular Engineering, Vanderbilt University, Nashville, Tennessee 37235, United States}
\alsoaffiliation{Vanderbilt Multiscale Modeling and Simulation (MuMS) Center, Vanderbilt University, Nashville, Tennessee 37235, USA}

\author{Janos Sallai}
\affiliation{Institute for Software Integrated Systems, Vanderbilt University, Nashville, Tennessee 37235, United States}

\author{Christopher R. Iacovella}
\email{christopher.r.iacovella@vanderbilt.edu}
\affiliation{Department of Chemical and Biomolecular Engineering, Vanderbilt University, Nashville, Tennessee 37235, United States}
\alsoaffiliation{Vanderbilt Multiscale Modeling and Simulation (MuMS) Center, Vanderbilt University, Nashville, Tennessee 37235, USA}


\begin{document}

\begin{abstract}
A key component to enhancing reproducibility in the molecular simulation community is reducing ambiguity in the parameterization of molecular models. Ambiguity in molecular models often stems from the dissemination of molecular force fields in a format that is not directly usable or is ambiguously documented via a non-machine readable mechanism.
Specifically, the lack of a general tool for performing automated atom-typing under the rules of a particular force field facilitates errors in model parameterization that may go unnoticed if other researchers are unable reproduce this process.
Here, we present Foyer, a Python tool that enables users to define force field atom-typing rules in a format that is both machine- and human-readable thus eliminating ambiguity in atom-typing and additionally providing a framework for force field dissemination.
Foyer defines force fields in an XML format, where SMARTS strings are used to define the chemical context of a particular atom type.
Herein we describe the underlying methodology of the Foyer package, highlighting its advantages over typical atom-typing approaches and demonstrate is application in several use-cases.
\end{abstract}

\section{Introduction}

Considerable efforts have been undertaken by many research groups to develop accurate classical force fields for a wide range of systems.\cite{Weiner1984,MacKerell2000,Jorgensen1996,siepmann1993simulating,Potoff2001,Sun1998,Oostenbrink2004}
Force fields are often expressed as a set of analytical functions with adjustable fitting parameters that describe the interactions between constituents of a system (often discrete atoms but, more generally, interaction sites).
Classical force fields are able to achieve high accuracy by creating sets of highly specific fitting parameters (i.e., atom types), in which each atom type describes an interaction site within a different chemical context.
The chemical context is typically defined by the bonded environment of an interaction site (e.g., the number of bonds and the identity of the bonded neighbors) and may also consider, among other factors, the bonded environment of the neighbors, and/or the specific molecule/structure within which the interaction site is included.
Consequently, a force field may include tens or even hundreds of different atom types for a given element.
For example, there are 347 different types of carbon in the OPLS forcefield parameter set distributed with GROMACS\cite{GromacsOPLS} where each type corresponds to a carbon atom within a different chemical context (i.e., a different atom type).
Thus, while force field development efforts have reduced -- or in some cases completely eliminated -- the need for researchers to generate their own fitting parameters,  determining \textit{which} parameters (i.e., atom types) to use can still be a tedious and error prone task.
Failure to properly identify the chemical context and atom type of an interaction site will inevitably lead to the incorrect implementation of the force field and thus inconsistent results.

Part of the difficulty in performing atom-typing (i.e., determining which atom type applies to an interaction site) stems from the fact that there is not yet a standardized way of unambiguously expressing chemical context and parameter usage.
As such, journal articles that report novel force field parameters may vary significantly in terms of their clarity.
In many cases, parameters are reported in a tabular format with minimal annotations and few (if any) examples of how to appropriately assign the atom types.
Since this approach does not allow for automated evaluation, different users of the force field may apply the atom types differently based on their own interpretation of the information provided.
Journal articles that utilize existing force fields often do not report the specific parameters used and typically do not specify which atom types were chosen for the interaction sites, instead providing citation(s) to the source of the force field parameters.
Even if the source of the parameters is clearly and fully specified, usage may  again depend on the clarity of the original source(s) and the interpretation by the end user, hampering reproducibility.
Force field parameter files that aggregate a large number of atom types (often thousands) into a single source suffer from some of the same issues.
 Often, they include only brief, unstructured  -- and sometimes ambiguous -- annotations as to parameter usage, and may, or may not, provide clear citations of the original source of the parameters.

To apply force fields, users can perform atom-typing manually (e.g., through the creation of an atom-typed template of a molecule), although manual assignment of parameters becomes tedious and error prone for large molecules and/or complex systems.  
Furthermore, manual manipulation of files is not considered a good practice in terms of reproducibility \cite{sandve2013ten} and manual assignment of parameters does not lend itself well to workflows such as screening, where thousands of unique systems with different chemical constituents and structures may need to be atom-typed.
To avoid manual assignment, end-users often develop in-house software to apply force fields in an automated fashion, however, such software is not typically made freely available to the broader community.
Without access to the same software, the exact atom-typing cannot be reproduced by others and if the source code is not made freely available, the logic used to interpret and apply the force field is unknown.
Similarly, if there are errors in the software/logic, these cannot be identified.
There exist a number of freely available atom-typing tools that read in a force field parameter file and execute a set of rules to apply the force field to a chemical topology \cite{Bush1993, Schuttelkopf2004a, Wang2006, Ribeiro2008, Malde2011, Vanommeslaeghe2012, Yesselman2012}, enabling the exact atom-typing process to be reproduced.
However, many of these atom-typing tools are either closed-source\cite{Schuttelkopf2004a, Malde2011}, simulation engine-specific\cite{Ribeiro2008, Yesselman2012}, and/or force field-specific\cite{Wang2006, Malde2011, Vanommeslaeghe2012}, which limits their utility.
Furthermore, these tools almost universally rely on a rigid hierarchy of rules\cite{Bush1993}, where rules must be called in a precise order such that more general atom types are only chosen when more specialized matches do not exist ($i.e.$, the order of rules defines the precedence).
Maintaining, let alone constructing, these hierarchies is challenging, especially for a large number of atom types.
In order to add a new atom type or correct an error in hierarchical schemes, a developer must have a complete picture of the hierarchy and know exactly where the relevant rule should be placed such that it does not inadvertently override other rules.
This may impose practical limits on functionality, where, for example, a user is not able to easily extend the rules to include new atom types, or that such attempts to extend the rules result in incorrect atom-typing for other systems.
For many tools, this approach is further complicated by the encoding of the hierarchy as a set of heavily nested if/else statements within the source code of the software.
These hierarchies may be difficult to validate and debug, and any changes or extensions to the rules, no matter how trivial, require modification of the source code itself.
Reproducibility issues may therefore arise if users make modifications or extensions to a piece of software and these changes are not made freely available to the larger community and/or incorporated into the main software distribution.
This also creates a situation where there are effectively two sets of rules since there is no guarantee that the logic statements in the source code (i.e., the machine readable rules) agree with the textual annotations in the force field parameter file (i.e., the human readable rules).

Several atom-typing tools have been developed that remove the need to encode atom type usage rules within the source code itself.
A unifying feature of these tools is the use of the simplified molecular-input line-entry system (SMILES \cite{weiniger1988smiles}), a language for describing chemical structures, or variants thereof, to define the chemical context of an atom type.
For example, Yesselman, et al. \cite{Yesselman2012} developed an atom-typing toolset for the CHARMM simulation engine, termed MATCH, that relies on assigning parameters by representing a molecule of interest as a graph and performing subgraph matching against a library of fragments with known parameters.
These fragments are represented as ``super smiles'', an extension of the SMILES language.
By using super smiles and storing these fragments in text files separate from the software, chemical context is expressed without the need to define a rigid if/else hierarchy within the software and thus new rules (i.e., fragments encoded as super smiles) can be added without modifying the code used to evaluate them, although,  the order of these fragments will still determine the final atom-type.
In other work, the Enhanced Monte Carlo (EMC) software developed by in't Veld \cite{emc} encodes chemical context of an atom type using SMILES, allowing usage rules to be defined outside of the source code, but EMC also relies on a hierarchical approach to set precedence.
In recent work, Mobley and coworkers\cite{Mobley2018} have developed the SMIRNOFF specification that effectively eliminates explicit atom types altogether, instead using SMIRKS (another language related to SMILES \cite{DaylightSMARTS}) to identify chemical fragments that are associated with a set of force field parameters, including bonded interactions. 
Similar to Yesselman, et al. \cite{Yesselman2012}, SMIRNOFF relies on representing the system as a graph and rules as subgraphs; in this approach rules must also be defined in a specific order to appropriately define precedence.
In all cases, the use of the SMILES-based approaches not only removes  the need to encode usage within the source code, but associates parameters with a human and machine readable definition of their chemical context.
However, these approaches still require rules be specified in the particular order to enable correct atom-typing and since the SMILES language is only specified for atomistic system, such approaches may have limited compatibility with, e.g., coarse-grained force fields.  

In this work, we present |Foyer|\cite{foyer}, a Python library for performing atom-typing based upon first-order logic over graph structures, designed to address many of the aforementioned issues associated with encoding chemical context, performing atom-typing of both atomistic and non-atomistic systems, and disseminating force fields to the community.
|Foyer| relies upon a force-field-agnostic formalism to express atom-typing and parameterization rules in a way that is expressive enough to be human readable while simultaneously being machine readable, allowing a single, unambiguous format to be constructed, suitable for use both the software and for dissemination. 
This logic is implemented via SMARTS\cite{DaylightSMARTS} to encode chemical context and ``|overrides|'' statements to define rule precedence.
SMARTS is an extension of the SMILES language that supports substructure definitions and allows expression of greater chemical detail and logic operations within the chemical patterns.
By using SMARTS to define chemical context,  atom type definitions are encoded outside of the source code, and thus force fields can be created and evolved without modification to the code used to evaluate them.
|Foyer| supports both fully atomistic and non-atomistic force fields, such as coase-grained and united-atom force fields, via an extension of SMARTS that enables user-defined ``elements'' (not in the periodic table) to be leveraged within the chemical context definitions.
Similar to Refs. \citenum{Yesselman2012, Mobley2018}, |Foyer| treats the system to be atom-typed as a graph and atom types are determined by identifying which subgraphs (specified via SMARTS) match the environment of a node in the graph (i.e., the chemical context of an interaction site).
In |Foyer|, the method of resolving atom types differs from most tools in that it is iterative and does not rely on rigid rule hierarchies.
Rule precedence is explicitly defined by the aforementioned |overrides| statements, thus atom-typing rules can appear in any order in the file, providing increased flexibility and eliminating physical placement in the file as a source of error.
Since this iterative approach evaluates all rules (not just the first to evaluate to "True"), automated evaluation can be used to help ensure that |Foyer| force field definitions (1) encompass all atom types in the force field and (2) are sufficiently descriptive without conflicting rules, both necessary conditions for publishing force fields in a way that is unambiguous and reproducible.
|Foyer| uses an XML-formatted force file format to define force field parameters and how to apply them to chemical systems.
Rather than creating a new file format, |Foyer| builds upon the OpenMM force field file format, extended to support the definition of the associated SMARTS definitions, |overrides| statements, and textual descriptions of the parameters; this format is also extended to support the definition of digital objective identifiers (DOIs) for each atom type, such that the source of the parameters can be clearly identified. 
By using XML, this force field file format is flexible, extensible, and by the nature of XML itself, provides clear labels/metadata of the contents in the file, improving human readability and therefore reducing ambiguity.
The |Foyer| software provides routines that create syntactically correct input files for a variety of common simulation engines and is designed to take, as input, chemical topologies from several of other community developed tools (e.g., |ParmEd|\cite{parmed}, |OpenMM|\cite{Eastman2013}, and |mBuild|\cite{Sallai2014, Klein2016,mBuild}).
We demonstrate that, by combining this software and annotation scheme with version control, force fields can be created and evolved in a transparent manner, allowing force field parameters and their usage to be more easily disseminated to the community.
Furthermore, since the force field format is both human and machine readable, force fields can be automatically validated to ensure completeness, allowing force field parameters and their usage to be published with reduced ambiguity, helping to improve reproducibility of molecular simulations.

\section{Defining chemical context and rule precedence}
\subsection{XML File Format}
|Foyer| utilizes the OpenMM\cite{Eastman2013} force field XML format to encode parameters, where this format is extended to allow for the definitions of chemical context and rule precedence (discussed below).  
To briefly summarize the OpenMM file format, atom types and forces are encoded as XML tags with various attributes defining the types of elements that they apply to (by name only), as well as the associated parameters for that interaction (e.g., the equilibrium bond length and spring constant for a harmonic bond).
Listing \ref{lst:xml_format} provides an example of encoding the OPLS force field parameters for linear alkanes in the OpenMM XML format (note, this Listing does not include our extensions).
As shown in Listing \ref{lst:xml_format}, the XML format provides clear descriptions of each of the parameters/properties defined in the file (e.g., |element="C"| indicates the entry is defining a carbon atom), along with additional tags that provide unambiguous descriptions of the types of interactions being used (e.g., the |<HarmonicBondForce>| tag is used to define the use of a harmonic force to define bonds). 
As such, this file format includes a wealth of metadata that is both human and machine readable.  
For more detailed information, we refer the reader to the OpenMM manual where this force field file format is extensively documented  (\url{http://docs.openmm.org/7.0.0/userguide/application.html#writing-the-xml-file}).

The flexible nature of XML allows it to be readily extended via the addition of new tags/attributes without fundamentally changing the original format, as new tags/attributes can simply be ignored by software that does not require them.
As shown in Table \ref{table:attributes}, and discussed in detail later, four new attributes have been added to the atom type entries in the existing OpenMM XML file format to enable the functionality needed to encode usage rules in |Foyer|: |def|, |desc|, |doi|, and |overrides|. The use of XML additionally allows sanity checks to be performed by using  XML schemas to ensure the expected attributes have been provided in the file.  

\begin{minipage}{\linewidth}
\begin{lstlisting}[caption={OpenMM formatted XML file for linear alkanes using the OPLS force field.  }, label=lst:xml_format, language=XML, frame=tlrb]
<ForceField>
  <AtomTypes>
    <Type name="opls_135" class="CT" element="C" mass="12.01100"/>
    <Type name="opls_136" class="CT" element="C" mass="12.01100"/>
    <Type name="opls_140" class="HC" element="H" mass="1.00800"/>
  </AtomTypes>
  <HarmonicBondForce>
    <Bond class1="CT" class2="CT" length="0.1529" k="224262.4"/>
    <Bond class1="CT" class2="HC" length="0.1090" k="284512.0"/>
  </HarmonicBondForce>
  <HarmonicAngleForce>
    <Angle class1="CT" class2="CT" class3="CT" angle="1.966986067" k="488.273"/>
    <Angle class1="CT" class2="CT" class3="HC" angle="1.932079482" k="313.800"/>
    <Angle class1="HC" class2="CT" class3="HC" angle="1.881464934" k="276.144"/>
  </HarmonicAngleForce>
  <RBTorsionForce>
    <Proper class1="CT" class2="CT" class3="CT" class4="CT" c0="2.9288"\\
        c1="-1.4644" c2="0.2092" c3="-1.6736" c4="0.0" c5="0.0"/>
    <Proper class1="CT" class2="CT" class3="CT" class4="HC" c0="0.6276"\\
        c1="1.8828" c2="0.0" c3="-2.5104" c4="0.0" c5="0.0"/>
    <Proper class1="HC" class2="CT" class3="CT" class4="HC" c0="0.6276"\\
        c1="1.8828" c2="0.0" c3="-2.5104" c4="0.0" c5="0.0"/>
  </RBTorsionForce>
  <NonbondedForce coulomb14scale="0.5" lj14scale="0.5">
    <Atom type="opls_135" charge="-0.18" sigma="0.35" epsilon="0.276144"/>
    <Atom type="opls_136" charge="-0.12" sigma="0.35" epsilon="0.276144"/>
    <Atom type="opls_140" charge="0.06" sigma="0.25" epsilon="0.12552"/>
  </NonbondedForce>
</ForceField>
\end{lstlisting}
\end{minipage}

\begin{table}
\caption{Extensions to the atom type definitions in the OpenMM XML format.} 
\label{table:attributes}
\begin{tabular}{ l  l  c }
\hline
\textbf{Attribute} & \textbf{Description} & \textbf{Example} \\ \hline
|def| & Defines the chemical context of an atom type via SMARTS & |[C;X4](H)(H)(H)C| \\ \hline
|desc| & Textual description of the atom type & |Alkane CH3| \\ \hline
|doi| & Digital object identifier to the atom type source & |10.1021/ja9621760| \\ \hline
|overrides| & Atom type(s) the current rule is given precedence over & |opls_136| \\ \hline
\end{tabular}
\end{table}

\UndefineShortVerb{\|}
\DefineShortVerb{\+}
\begin{table}
\small
\caption {2D depictions of molecular fragments referred to in the text} \label{tab:chemdraw}
\begin{tabular}{|c c c|c c c|c c|}
\hline
\multicolumn{3}{|c|}{\textbf{Alkane}} & \multicolumn{3}{c|}{\textbf{Alkene}} & \multicolumn{2}{c|}{\textbf{Benzene}} \\ \hline
\textbf{C, CH$_3$} & \textbf{C, CH$_2$} & \textbf{H} & \textbf{C, (R2-C=)} & \textbf{C, (RH-C=)} & \textbf{H} & \textbf{C} & \textbf{H} \\ \hline
\includegraphics[width=0.1\textwidth]{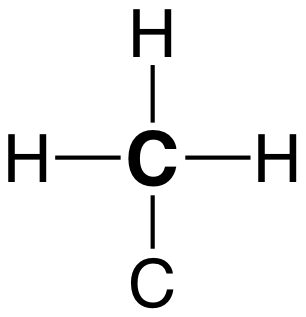}
& \includegraphics[width=0.1\textwidth]{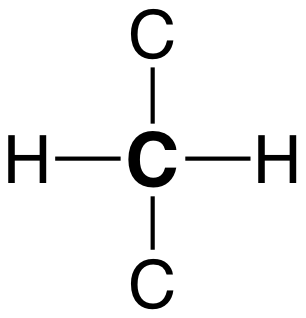}
& \includegraphics[width=0.1\textwidth]{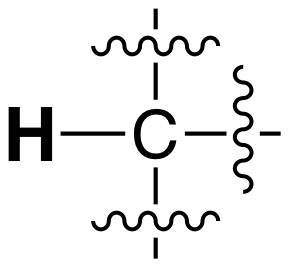}
& \includegraphics[width=0.1\textwidth]{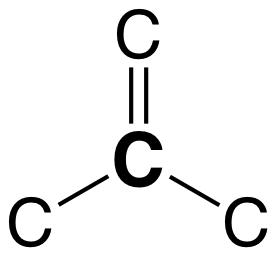}
& \includegraphics[width=0.1\textwidth]{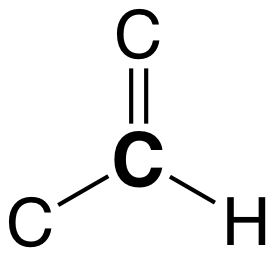}
& \includegraphics[width=0.1\textwidth]{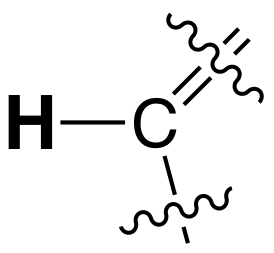}
& \includegraphics[width=0.1\textwidth]{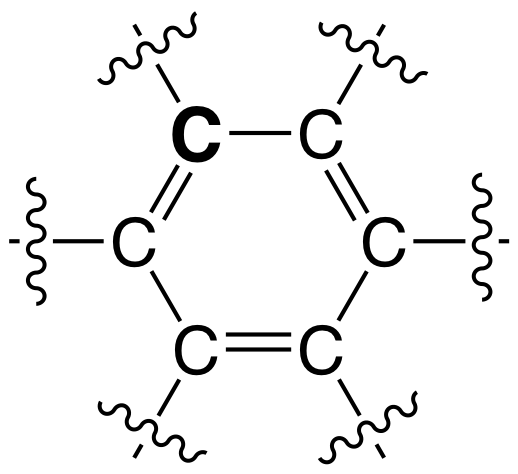}
& \includegraphics[width=0.1\textwidth]{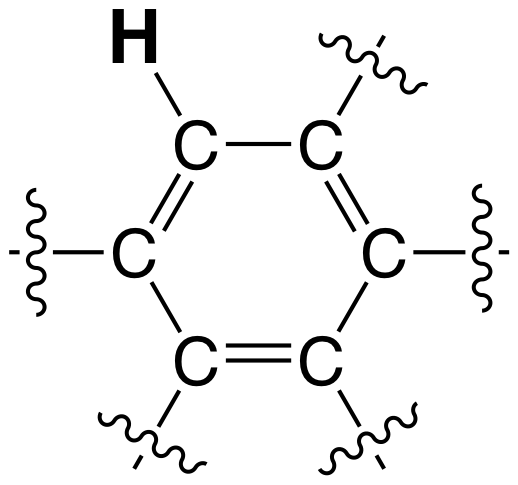} \\
+opls_135+ & +opls_136+ & +opls_140+ & +opls_141+ & +opls_142+ & +opls_144+& +opls_145+ & +opls_146+ \\ \hline
\end{tabular}
\end{table}
\UndefineShortVerb{\+}
\DefineShortVerb{\|}

\subsection{Using SMARTS to define chemical context}

\begin{table}[]
\begin{threeparttable}
\caption {Currently implemented SMARTS atomic primitives$^a$} \label{tab:smarts}
\begin{tabular}{ l l l l }
\hline
\textbf{Symbol} & \textbf{Symbol name} & \textbf{Atomic property requirements} & \textbf{Default} \\ \hline
* & wildcard & any atom & (no default) \\ \hline
A & aliphatic & aliphatic & (no default) \\ \hline
r\textless{}n\textgreater{} & ring size & in smallest SSSR$^b$ ring of size \textless{}n\textgreater{} & any ring atom \\ \hline
X\textless{}n\textgreater{} & connectivity & \textless{}n\textgreater total connections & exactly one \\ \hline
\#n & atomic number & atomic number \textless{}n\textgreater{} & (no default) \\ \hline
\end{tabular}
\begin{tablenotes}
\item $^a$This table has been adapted from the Daylight SMARTS website. \cite{DaylightSMARTS}
\item $^b$Smallest set of smallest rings.
\end{tablenotes}
\end{threeparttable}
\end{table}

\begin{table}[]
\begin{threeparttable}
\caption {Extensions to SMARTS atomic primitives developed in this work} \label{tab:smarts-ext}
\begin{tabular}{ l l l l }
\hline
\textbf{Symbol} & \textbf{Symbol name} & \textbf{Atomic property requirements} & \textbf{Default} \\ \hline
\_A & non-element & non-atomistic element & (no default) \\ \hline
\%\textless{}type\textgreater{} & atomtype & of atomtype \textless{}type\textgreater{} & (no default) \\ \hline
\end{tabular}
\end{threeparttable}
\end{table}

\begin{table}[]
\begin{threeparttable}
\caption {SMARTS Logical Operators$^a$} \label{tab:smarts-op}
\begin{tabular}{ l l l }
\hline
\textbf{Symbol} & \textbf{Expression} & \textbf{Meaning} \\ \hline
exclamation & !e1 & not e1 \\ \hline
ampersand & e1\&e2 & e1 and e2 (high precedence) \\ \hline
comma & e1,e2 & e1 or e2 \\ \hline
semicolon & e1;e2 & e1 and e2 (low precedence) \\ \hline
\end{tabular}
\begin{tablenotes}
\item $^a$This table has been adapted from the Daylight SMARTS website. \cite{DaylightSMARTS}
\end{tablenotes}
\end{threeparttable}
\end{table}

The chemical context of an interaction site is typically defined by its bonded environment, notably the number of bonds and the identities of bonded neighboring interaction sites, but may also  include longer range information, such as the bonded environment of neighbors.
To encode this information, |Foyer| utilizes SMARTS\cite{smarts}, a language for defining chemical patterns. 
SMARTS is an extension of the more commonly used SMILES\cite{weininger1988smiles} notation, providing additional tokens that enable users to express greater chemical detail and logic operations.
SMARTS notation is expressed as strings that simultaneously include arbitrary chemical complexity but are concise and clear enough for human consumption, in addition to being machine readable.
As an example, consider defining the chemical context of OPLS-AA atom types for carbon and hydrogen atoms in a linear alkane, as shown in Listing \ref{lst:methyl_atomtypes} (note, only the |<AtomTypes>| section of the file is shown, as this is the only section that differs from Listing \ref{lst:xml_format}).
The reader is referred to Table \ref{tab:chemdraw} for a visual depiction of these atom types.
To encode the chemical context, the |def| attribute is added to the OpenMM XML format to encode the corresponding SMARTS string.
Here, the atom type that specifies the terminal methyl group, |opls_135|   (``-CH3'') can be expressed as |[C;X4](C)(H)(H)H| in the SMARTS notation.
In this SMARTS notation, |[C;X4]| indicates that the element of interest -- always the first token in the SMARTS string -- is a carbon atom (i.e., |C|) and this carbon atom  has 4 total bonds (i.e., |;X4|, where |;| indicates the logical operator |AND|). 
 The identities of the 4 bonded neighbors are 1 carbon atom and 3 hydrogen atoms, expressed as |(C)(H)(H)H|.
Similarly, the |opls_136| atom type, which describes a methylene group in an alkane, is expressed in SMARTS notation as |[C;X4](C)(C)(H)H|.
Here, the only change from the |opls_135| SMARTS definition lies in the identity of the 4 bonded neighbors (2 carbon atoms and 2 hydrogen atoms).
Increased chemical complexity can be described by adding details about each of neighboring interaction sites within SMARTS.
For example, the |opls_140| atom type, which describes a generic alkane hydrogen, is defined as |H[C;X4]| - a hydrogen atom bonded to a carbon atom with 4 bonds.
Multiple valid SMARTS can be defined for each atom type, where, e.g., |opls_140| could be defined simply as |def="H"| since there is only a single hydrogen atom type defined in Listing \ref{lst:methyl_atomtypes}, although we note that such a simplified definition would not necessarily provide a user of the force field with a clear understanding of the chemical context for which this atom type applies.
The |Foyer| extensions to the OpenMM XML file format also include the |desc| attribute (e.g., shown in Listing \ref{lst:methyl_atomtypes}) that allows for unstructured comments to be provided in addition to the SMARTS definitions.

We note that the parser in the |Foyer| libraries does not currently support the full SMARTS language, instead providing support for the subset that was found to be relevant to the definition of chemical context for atom types. Table \ref{tab:smarts} lists the currently supported primitives, Table \ref{tab:smarts-ext} shows our extensions to the language, and Table \ref{tab:smarts-op} outlines the logical operators supported.  Increased support for the SMARTS language will be provided in future software releases, as needed. 

\begin{minipage}{\linewidth}
\begin{lstlisting}[caption={Atom type definitions for carbon and hydrogen atoms in a linear alkane using the OPLS force field implemented in Foyer.  Note, only the section that applies to atom types is shown for clarity.}, label=lst:methyl_atomtypes, language=xml, frame=tlrb]
<ForceField>
  <AtomTypes>
    <Type name="opls_135" class="CT" element="C" mass="12.01100" \\
           def="[C;X4](C)(H)(H)H" desc="alkane CH3"/>
    <Type name="opls_136" class="CT" element="C" mass="12.01100"\\
           def="[C;X4](C)(C)(H)H" desc="alkane CH2"/>
    <Type name="opls_140" class="HC" element="H" mass="1.00800"\\
           def="H[C;X4]" desc="alkane H"/>
  </AtomTypes>
</ForceField>
\end{lstlisting}
\end{minipage}

\subsection{Establishing rule precedence}
Rule precedence must be established when multiple atom type definitions can apply to a given interaction site. 
In typical hierarchical schemes, this is determined implicitly by the order in which rules are evaluated; in general, more specific rules are evaluated first and when a match is found, the code stops evaluating rules altogether.
While this approach works, it becomes more challenging to maintain the correct ordering of rules as the number of atom types grows and as chemistries become more complex and specific.
Users may find it difficult, if not impossible, to make even small additions to a larger force field without breaking existing behavior.
|Foyer| allows rule precedence to be explicitly stated via the use of the |overrides| attribute added to the XML file format.
This allows atom type usage rules to be encoded in any order within the file, eliminating incorrectly placed rule order as a source of error.
|Foyer| iteratively evaluates all rules on all interaction sites in the system, maintaining for each  interaction site a ``whitelist'' consisting of rules that evaluate to |True| and a ``blacklist'' consisting  of rules that have been superseded by another rule (i.e., those that appear in the |overrides| attribute).
The set difference between the white- and blacklists of an interaction site yields the correct atom type if the force field is implemented correctly (incorrect/incomplete definition of force fields is discussed later).
As an example of a system where |overrides| need to be defined, consider describing alkenes and benzene in a single force field file, as shown in Listing \ref{lst:overrides} (note, again only the |<AtomTypes>| section of the force field file is shown).
The reader is again referred to Table \ref{tab:chemdraw} for visual depictions of the relevant atom types.

\begin{minipage}{\linewidth}
\begin{lstlisting}[caption={Atom type definitions for alkenes and benzene using the OPLS force field as implemented in Foyer, highlighting the overrides syntax and mechanism for referencing other atom types. Note, only the section that applies to atom types is shown for clarity.}, label=lst:overrides, language=xml, frame=tlrb]
<ForceField>
  <AtomTypes>
    <Type name="opls_141" class="CM" element="C" mass="12.01100" \\
           def="[C;X3](C)(C)C" desc="alkene C (R2-C=)"/>
    <Type name="opls_142" class="CM" element="C" mass="12.01100" \\
           def="[C;X3](C)(C)H" desc="alkene C (RH-C=)"/>
    <Type name="opls_144" class="HC" element="H" mass="1.00800" \\
           def="[H][C;X3]" desc="alkene H"/>
    <Type name="opls_145" class="CA" element="C" mass="12.01100" \\
           def="[C;X3;r6]1[C;X3;r6][C;X3;r6][C;X3;r6][C;X3;r6][C;X3;r6]1" \\
           overrides="opls_142" desc="benzene C"/>
    <Type name="opls_146" class="HA" element="H" mass="1.00800" \\
           def="[H][C;%opls_145]" overrides="opls_144" desc="benzene H"/>
  </AtomTypes>
</ForceField>
\end{lstlisting}
\end{minipage}

When atom-typing a benzene molecule, the carbon atoms in the ring will match the SMARTS patterns for both |opls_142| (an alkene carbon) and |opls_145| (a benzene carbon).
Without the |overrides| attribute, |Foyer| will find that multiple atom types apply to each carbon atom.
Providing the |overrides| attribute indicates that if the |opls_145| pattern matches, it will supersede |opls_142|.
Thus, the difference between the whitelist (containing |opls_142| and |opls_145|) and blacklist (containing only |opls_142|) would be |opls_145|.

Note that multiple atom types can be listed in a single |overrides| attribute, however, the approach taken here also allows atom types to inherit |overrides| from the atom types they override.
For example, consider a case in which atom types 1, 2, and 3 each evaluate to |True| for an interaction site.
If atom type 3 overrides atom type 2 (i.e., adds atom type 2 to the blacklist) and atom type 2 overrides atom type 1 (i.e., adds atom type 1 to the blacklist), then atom type 3 will implicitly override atom type 1.

Additionally, in |Foyer|, the SMARTS grammar has been modified such that specific atom type names can also be included within the definition (see Table \ref{tab:smarts-ext}).
For example, |opls_146|, the hydrogen atom attached to carbon atoms in a benzene ring, has the SMARTS definition |[H][C;
This states that the interaction site of interest is a hydrogen atom (|H|) and is bonded to a carbon atom that has atom type |opls_145| (|C;
Because |Foyer| evaluates rules iteratively for each interaction site, such recursive definitions can be utilized without the need to explicitly define atom types in a chemical topology input file.
For example, in this case, when |Foyer| identifies the interaction site of a carbon atom to be |opls_145|, the next iteration to evaluate the hydrogen atom will find that |opls_146| now evaluates to |True|.  
Similar to how an |overrides| statement clearly defines precedence, this recursive definition provides a clear way to identify chemical context and the relationship between different atom types for highly specific parameters. 
We note, that one could also replace the recursive reference to |opls_145|  with its SMARTS string,  although in this case it would result in a more complex,  less human readable definition. 

 
Because the logic used to define chemical context is separated from the source code used to evaluate it, one can construct a force field file that contains only the relevant subset of atom types need for a given application area.
Using the above example of benzene and alkenes, if a system only contained benzene molecules, one could avoid specifying the |overrides| attributes altogether by simply creating a force field file containing only atom types relevant to benzene and eliminating those associated with alkenes.
In many cases, considering smaller subsets is beneficial as the amount of effort required to differentiate and set rule precedence between atom types is reduced.
Additionally, using smaller files reduces the likelihood of errors related to defining chemical context and rule precedence, reduces the number of test molecules with known atom types required to fully validate the rules, and increases the readability of the force field files by limiting the number of entries.

\subsection{Extension of SMARTS for non-atomistic systems}
|Foyer| is able to atom-type systems in which an interaction site does not represent a single atom with a standard element, but instead may represent a group of atoms (relevant to united-atom and coarse-grained force fields) or a generic site (relevant to simplified models).
Standard SMARTS notation does not support non-atomic species due to its reliance on the presence of an element specification for each interaction site.
To circumvent this limitation, the |Foyer| SMARTS parser allows users to define custom ``elements'' by prefixing their string representation with an underscore (see Table \ref{tab:smarts-ext}).
For example, |_CCC| could represent a coarse-grained interaction site intended to model three carbon atoms.
In its current implementation, |Foyer| makes a first pass through force field files to detect any custom element definitions.
These are injected into the grammar that parses SMARTS strings and are given priority over standard elements.
This allows non-atomistic and atomistic atom types to be used, either separately or together. 

In practice, united-atom and coarse-grained force fields can be defined in an almost identical fashion to all-atom force fields, where the only difference is that ``elements'' are user-defined strings prepended with an underscore.
As an example, consider an alkane modeled with the united-atom TraPPE force field.
An interaction site in this force field represents both carbon and the hydrogen atoms bonded to it.
Thus, this force field contains two distinct atom types, one that represents $\mathrm{CH_3}$ (|_CH3|) and one that represents $\mathrm{CH_2}$ (|_CH2|).
These can be encoded as shown in Listing \ref{lst:alkane_trappe}.
Focusing on atom type |CH3_sp3|, usage is encoded with the  definition |[_CH3;X1][_CH3,_CH2]| which states that the base ``element'' is |_CH3|  with one bond (i.e., |;X1|) to either  a |_CH3| or a |_CH2| group.
In SMARTS, a comma indicates an ``OR'' logic  statement while a semicolon is used to denote an ``AND'' logical statement (see Table \ref{tab:smarts-op} for a complete list of SMARTS logical operators).
In this example, |[_CH3;X1]|  states the element must be |_CH3| ``AND'' have only a single bond.
Atom-type |_CH2_sp3|, which represents a ``middle'' alkane carbon and its two associated hydrogen atoms, is defined similarly as |[_CH2;X2]([_CH3,_CH2])[_CH3,_CH2]|.
Here, the base ``element'' is a |_CH2| with two bonds (i.e., |;X2|), each of which may be either ``element'' |_CH3| or |_CH2|.

\begin{minipage}{\linewidth}
\begin{lstlisting}[caption={Atom:type definitions for united atom alkanes using TraPPE.}, label=lst:alkane_trappe, language=xml, frame=tlrb]
<ForceField>
 <AtomTypes>
    <Type name="CH3_sp3" class="CH3" element="_CH3" mass="15.03500" \\
           def="[_CH3;X1][_CH3,_CH2]" desc="Alkane CH3, united atom"/>
  <Type name="CH2_sp3" class="CH2" element="_CH2" mass="14.02700" \\
           def="[_CH2;X2]([_CH3,_CH2])[_CH3,_CH2]" desc="Alkane CH2, united atom"/>
 </AtomTypes>
</ForceField
\end{lstlisting}
\end{minipage}

\subsection{Determining bonded parameters}

Once a chemical topology is atom-typed, bonded interactions can be determined by simply searching for the matching pairs, triplets, and quartets (bonds, angles, and torsions, respectively).
In many force fields, the bonded parameters are not as specific as the non-bonded interactions, and thus are not defined directly based on atom types.
Thus, rather than atom types, a more general |class| identifier (sometimes referred to as the ``bond family'') is used to identify these interactions.
In Listing \ref{lst:pfa}, both |opls_136| and |opls_962| are part of the same |class| ``CT''.
Thus a bond between |opls_136| and |opls_962| would have the same parameters (defined as |class1="CT" class2="CT"| in Listing \ref{lst:pfa}) as a bond between |opls_136| and |opls_136| (also defined as |class1="CT" class2="CT"|).
However, this general approach breaks down for certain chemical topologies.
For example, while the atom types for carbon atoms in alkanes and perfluoralkanes are both of |class| ``|CT|'' and share the same bond and angle parameters for carbon atoms, they differ in terms of torsional parameters.
In order to handle this conflict, many codes require users to comment out the more general set of parameters or include statements within the code that accomplish the same task.
However, in this approach, one would not be able to atom-type  a system composed of a mixture alkane and perfluoroalkane molecules\cite{Morgado2013}, since only one set of parameters can be included simultaneously.
Note that while one could define a new |class| to, for example, differentiate between alkanes and perfluoroalkanes, this would also require defining all the interactions with existing relevant |class| entries and result in a force field file with many duplicate parameters sets that simply have different labels.
The OpenMM format allows bonded parameters to be defined using the |type| attribute, referring directly to the |name| attribute that stores the atom type, instead of the |class| attribute, which allows bonded interactions to be defined with increased specificity.
Additionally, mixed use of |type| and |class| in the definition of these bonded interactions is supported.
Referring  to Listing \ref{lst:pfa}, to provide the necessary distinction between torsional parameters for perfluoroalkanes and alkanes, one could define perfluoroalkane torsions using type attributes (i.e., |type1="opls_962" type2="opls_962" type3="opls_962" type4="opls_962"|, where |opls_962| is defined in the |<AtomTypes>| XML section), and alkane torsions with the more general quartet for alkanes of |class1="CT" class2="CT" class3="CT" class4="CT"| that uses |class| attributes.

When iterating through bonded parameter definitions, OpenMM assigns parameters based on the first match found.
In the example described above, perfluoroalkane torsional parameters would therefore need to be defined before alkane parameters in the torsional section, and thus the ordering shown in Listing \ref{lst:pfa} would result in the incorrect assignment of torsional parameters for perfluoroalkanes.
While |overrides| statements could be used to set rule precedence for bonded topologies and thus eliminate the need to specify order in the file, additional modification to the force field file format would be required because bonded parameters do not have a ``name'' attribute like atom types.
A more simple approach taken by |Foyer| is to perform a preprocessing step on the bonded parameters.
This step orders  bonded parameters such that the most specific cases are sorted to the top of the list to set precedence.
This is accomplished by assigning a weight to each entry proportional to the number of |type| attributes included.
For example, a torsion that defines the atom types for which it applies (i.e., has 4 |type| attributes) would be given the highest weight, whereas an entry that specifies only |class| attributes would be given the lowest.
Thus, for the force field XML shown in Listing \ref{lst:pfa} Foyer would reverse the order of the two defined dihedrals during preprocessing.

\begin{minipage}{\linewidth}
\begin{lstlisting}[caption={Foyer force field XML snippet showing atom types defined for carbon in CH$_2$ and CF$_2$ substructures, a bonded definition between carbons, and C-C-C-C dihedral definitions for hydrogenated and perfluorinated alkanes.}, label=lst:pfa, language=XML, frame=tlrb]
<ForceField>
  <AtomTypes>
   ...
    <Type name="opls_136" class="CT" element="C" mass="12.01100" \\
        def="[C;X4](C)(C)(H)H" desc="alkane CH2" />
    <Type name="opls_962" class="CT" element="C" mass="12.01100" \\
        def="[C;X4](C)(C)(F)F" desc="perfluoroalkane CF2" />
   ...
  </AtomTypes>
  <HarmonicBondForce>
   ...
    <Bond class1="CT" class2="CT" length="0.1529" k="224262.4"/>
   ...
  </HarmonicBondForce>
   ...
  <RBTorsionForce>
   ...
    <Proper class1="CT" class2="CT" class3="CT" class4="CT" c0="2.9288" \\
        c1="-1.4644" c2="0.2092" c3="-1.6736" c4="0.0" c5="0.0"/>
    <Proper type1="opls_962" type2="opls_962" type3="opls_962" type4="opls_962" \\
        c0="14.91596" c1="-22.564312" c2="-39.41328" c3="11.614784" \\
        c4="35.446848" c5="0.0"/>
   ...
  </RBTorsionForce>
   ...
</ForceField>
\end{lstlisting}
\end{minipage}

\section{Foyer software}
\begin{figure}[ht!]
    \centering
    \includegraphics[width=0.5\textwidth]{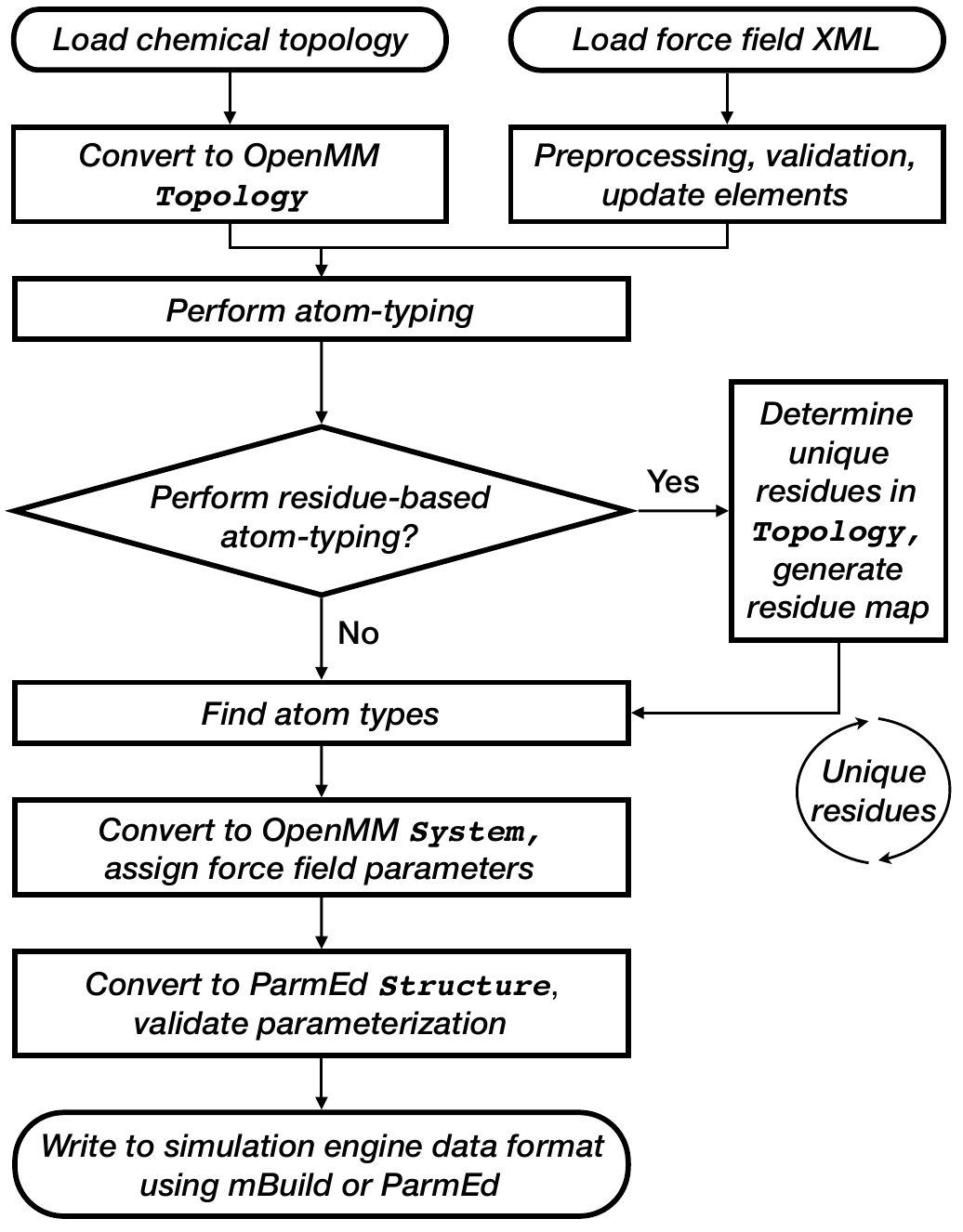}
    \caption{Flowchart of the Foyer software from chemical topology and force field XML inputs to a simulation data file output.}
    \label{figure:flowchart}
\end{figure}

The |Foyer| software has been developed as a Python library that can read the force field XML specification discussed above and perform atom-typing.
Python allows for portability between platforms and provides a wealth of freely available modules (e.g., NumPy \cite{Oliphant:2015:GN:2886196}, SciPy \cite{scipy}, NetworkX \cite{networkx}) to facilitate many of the underlying operations.
The source, documentation, and examples of |Foyer| usage are all freely available and can be found on the project's GitHub repository \url{https://github.com/mosdef-hub/foyer} and associated website \url{http://mosdef-hub.github.io/foyer/}.
Foyer is a core package within the Molecular Simulation Design Framework (MoSDeF), a framework/protocol aimed towards fully reproducible molecular simulations, and is hosted under the |mosdef-hub| GitHub organization.
Additionally, tutorials to Foyer's usage can be accessed via a separate tutorial repository \url{https://github.com/mosdef-hub/foyer_tutorials}.
Figure \ref{figure:flowchart}  provides an overview of the general software workflow, which will be discussed here.

\subsection{Inputs and Preprocessing}
|Foyer| accepts, as input, the XML force field file and a chemical topology for which to apply the force field.
In addition to sorting bonded parameters by specificity as described in the previous section, the XML force field file undergoes a preprocessing and validation step via application of an XML schema definition.
This schema definition in |Foyer| enforces which elements (e.g. |HarmonicBondForce|) are valid and how their attributes should be formatted.
While this does not test the accuracy of the parameters, it does ensure that all of the expected parameters are defined.
Additionally, the schema ensures that atom types are not defined more than once and that atom types referenced in other sections (e.g., |<HarmonicBondForce>|) are actually defined in the |<AtomTypes>| section.
Next, the SMARTS strings defined by the |def| attribute for all atom types are parsed and checked for validity.
This does not validate whether a SMARTS string is correctly defined for a given interaction site but simply ensures that the SMARTS string can be interpreted by |Foyer| and does not contain any erroneous characters.
Parsing errors are captured and re-raised with error messages that allow a user to pin point the location of the problem in the XML file and within the SMARTS string.
Wherever possible, |Foyer| attempts to provide helpful suggestions for fixing detected errors.

Input chemical topologies can be passed to |Foyer| through various data structures; the current version supports  the |OpenMM| |Topology| object\cite{}, the |ParmEd| |Structure| object\cite{parmed,shirts2017lessons}, and the |mBuild| |Compound| object\cite{Sallai2014,Klein2014}.
Each of the |OpenMM|, |ParmEd|, and |mBuild| topologies support inputs from a variety of common molecular file formats, such as PDB and MOL2, and thus it is typically straightforward to convert a given system into a data structure that |Foyer| can accept.
Regardless of the input format, once read into |Foyer| the chemical topology is converted to an |OpenMM| |Topology| object, in the current release of the software.
The |OpenMM| |Topology| object provides a standardized data container to store the necessary system information and allows for leveraging of routines already defined within |OpenMM|'s library.

\subsection{Atom-typing}
\begin{figure}[ht!]
    \centering
    \includegraphics[width=0.5\textwidth]{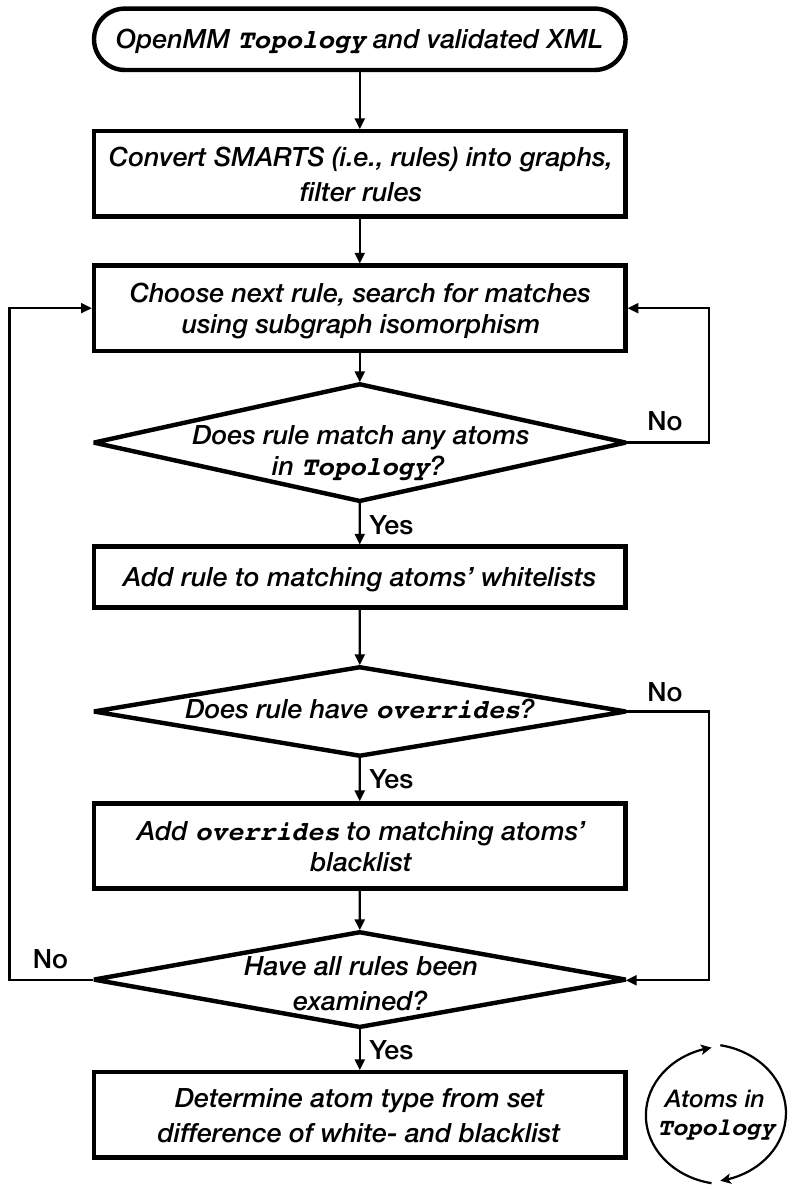}
    \caption{Flowchart of Foyer's atom-typing process.}
    \label{figure:flowchart-atomtype}
\end{figure}

A flowchart of |Foyer|'s atom-typing procedure is shown in Figure. \ref{figure:flowchart-atomtype}.
To perform atom-typing, |Foyer| constructs a graph of the complete system defined by the chemical topology (or alternatively a graph of each unique residue, see the Residue-based Atom-typing section below) and iteratively searches for SMARTS matches via subgraph isomorphism (where subgraphs are generated for each SMARTS definition).
Graph construction and matching are performed using the |NetworkX| package, an open-source Python project that provides an intuitive interface for a multitude of graph-based algorithms and is the de facto standard network analysis library in Python.
During this step, the iterative process of determining the atom type is undertaken, adding rules to the white and back lists for each interaction site in the system.

\begin{figure}[ht!]
    \centering
    \includegraphics[width=\textwidth]{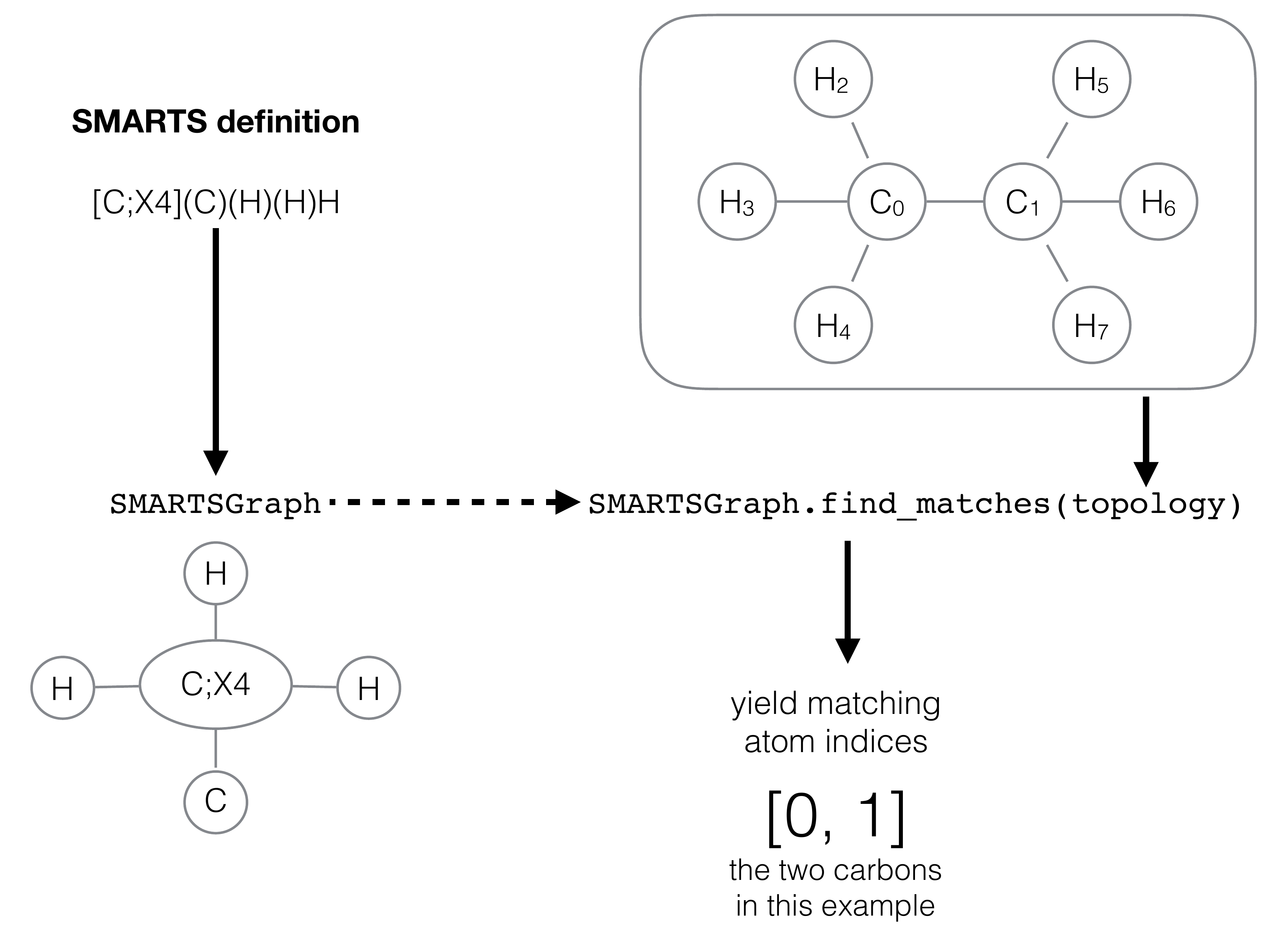}
    \caption[Schematic of the workflow to apply SMARTS patterns to chemical topologies.]
    {
Schematic of the workflow to apply SMARTS patterns to chemical topologies.
    The SMARTS strings used to define atomtypes are read into a \texttt{SMARTSGraph} class which inherits from \texttt{NetworkX}'s core data structure.
    Using the \texttt{find\_matches} method, a \texttt{SMARTSGraph} instance can search for subgraph isomorphisms of itself within a provided chemical topology and will yield all atoms that match the first token in the original SMARTS string.
    }
    \label{figure:subgraph}
\end{figure}

The implementation of the SMARTS based atom-typing scheme is comprised of several steps and internally relies on a subgraph isomorphism to detect matches as highlighted in Figure~\ref{figure:subgraph}.
First, a SMARTS string is parsed into an abstract syntax tree (AST) from which we populate a |SMARTSGraph| object.
This class inherits from the |Graph| class in the |NetworkX| package\cite{networkx}.
Elements in this |SMARTSGraph| are represented as nodes and chemical bonds as edges.
Inheriting from |NetworkX| is convenient in that it allows us to leverage most of the algorithms and visualization methods already implemented there.
The primary distinguishing feature of the |SMARTSGraph| is the set of methods that encode the logic for matching the more complex SMARTS tokens.
These methods can be directly used by |NetworkX|'s implementation of the VF2 subgraph isomorphism algorithm\cite{vf2}.
A thin wrapper provided by the |find_matches| method allows a |SMARTSGraph| instance to search for all subgraph isomorphisms within a bare chemical topology (an non-atom-typed graph of just elements and bonds).
This method returns the indices of all elements that match the first token in the SMARTS string, which defines the atom type that we are looking for.
Successfully matching elements have the atom type definition added to their whitelist and any overridden types added to their blacklist.
The appropriate atom type for an interaction site is determined by examining the difference between white- and blacklists, where a sufficiently descriptive force field should yield only a single atom type as the difference between the two lists. 

The use of white- and blacklists provides users with a means to validate the completeness of the chemical contexts defined by the set of SMARTS strings and |overrides|.
For example, when considering a given atom, if multiple valid atom types are found as the difference between that atom's white- and blacklists,  this indicates the rule definitions within the force field are not sufficiently unique and likely have incomplete information provided to the |override| attributes.
|Foyer| provides  the list of conflicting types to aid in resolving such issues.
If no atom types exist as the difference, the interaction site of interest cannot be described by the force field rules as implemented.
Typically, this will require adding a new atom type or amending an existing atom type's definition. 
This may also indicate,  that there is an error in how rule precedence has been defined, such as, all the rules on the whitelist ``overriding'' each other.
Note, the efficacy of this type of validation in |Foyer| will depend on providing a sufficient range of systems to fully explore the combinations of atom types that can be applied.
As a general rule, the set of systems chosen to perform validation tests should collectively utilize all atom types defined in the force field.

\subsubsection{Residue-based Atom-typing}
Many systems of interest to molecular simulation contain topologies that consist of duplicates of smaller molecules or repeat units, each with identical topologies.
A brute-force implementation of the atom-typing process wastes time by repeating subgraph isomorphism computation on each repeat unit.
Also, this approach does not scale well with system size.
To eliminate unnecessary calculations, a map of atom-typed residues is saved after each unique residue is atom-typed the first time.
Then, when an identical residue is found, it copies the atom-typed information from the residue map instead of repeating the subgraph isomorphism.
This feature is enabled by default but can optionally be turned off.

\begin{figure}
    \centering
    \includegraphics[width=0.5\textwidth]{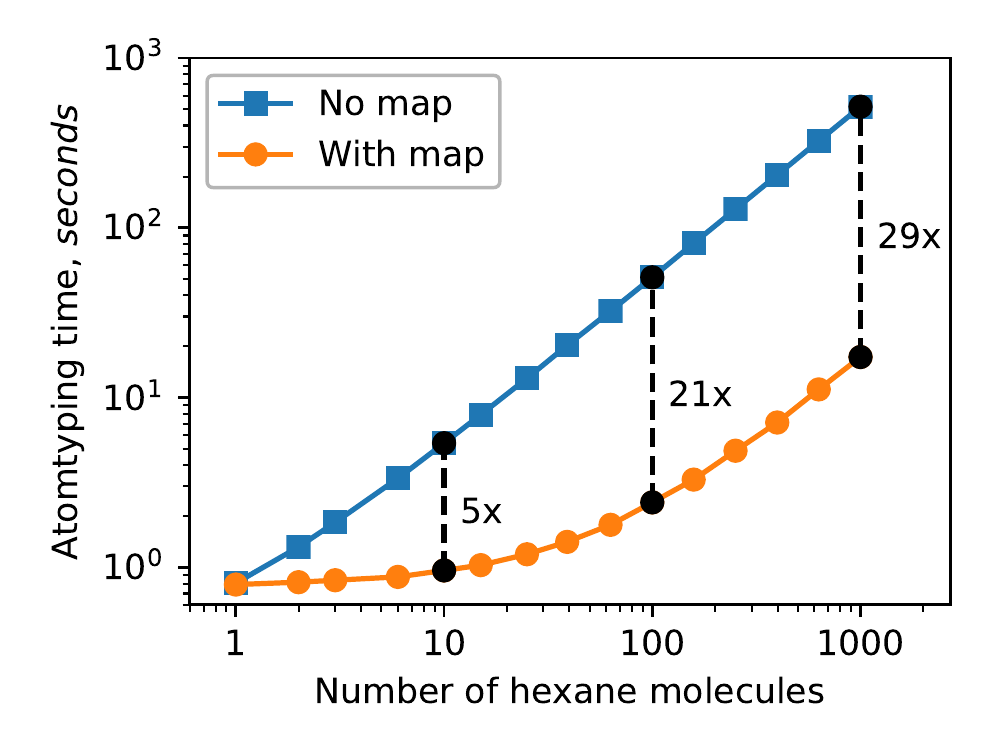}
    \caption[Caption]
    {
    Comparison of atom-typing cost with and without the use of a residue template.
    Without a residue template map, the scaling is approximately linear with system size.
    With a map, the scaling is independent of system size for small systems and becomes approximately linear at larger system sizes due to operations other than the subgraph isomorphism.
    The speedup approaches a factor of approximately 30 as the system size becomes large.
    The times to atom type each system were obtained with a 2013 MacBook Pro, 3GHz Core i7, 8 GB RAM.
    }
    \label{figure:residue}
\end{figure}

As a example, consider a box of $N$ hexane molecules.
After the subgraph isomorphism is called on the first molecule, the result is copied and saved into a map.
Then, when molecules 2 to $N$ are encountered, those results are copied into the running topology.
The time it takes for the |apply| function to finish is timed for each case and plotted in Figure~\ref{figure:residue} relative to the brute force approach; as can be seen, the use of the residue map speeds up the overall atom-typing step by more than an order of magnitude for common system sizes.

\subsection{Force field assignment and output}
Once atom-typed, the |OpenMM| |Topology| now contains atom types for all particles in the system and is used to create an |OpenMM| |System| object upon which bonded parameters of the system are determined.
This step can be accomplished by searching the list of bonded parameters for the appropriate pair, triplet, and quartet of atom types for bonds, angles, and dihedrals, respectively; such routines exist within OpenMM and are utilized in this context, where again we note these interactions undergo a sorting to ensure more specific definitions appear first in the file. 
Additionally, validation checks are performed at this time to ensure all triplets and quartets of interaction sites have had angle and dihedral (proper and improper) parameters assigned (checks for bond parameterization of interaction site pairs are performed by |OpenMM| in the prior step).
These validation checks provide the user with an error (that can optionally be overridden) to help prevent the return of incorrectly parameterized |Structures|.
To output the atom-typed system into a usable format for a simulation engine, the fully atom-typed and parameterized system is returned as a |ParmEd| |Structure| object.
Through the use of the |ParmEd| |Structure|, |Foyer| has access to various additional functionality, such as I/O routines that properly parse the |ParmEd| |Structure| into common chemical file formats (MOL2, PDB, NAMD and GROMACS formats, among others\cite{parmed}).
For file formats not natively supported by |ParmEd|, custom I/O routines for outputting to these formats (e.g., the LAMMPS data file format) have been developed within the |mBuild| package.

It should be noted that by utilizing |ParmEd|, the force fields that |Foyer| currently supports must match functional forms supported by the internals of the |ParmEd| |Structure| object.
For example, non-bonded interactions are currently limited to a 12-6 Lennard-Jones functional form.
However, due to the large amount of force fields that utilize this functional form (e.g., Amber\cite{Weiner1984}, GAFF\cite{Wang2004}, OPLS\cite{Jorgensen1996}, TraPPE\cite{Martin1998,trappe_website}, CHARMM\cite{MacKerell2000}), the current version of |Foyer| is still widely applicable; future development will include support for additional functional forms to better accommodate the diverse force field landscape that exists.

\subsection{Validating output}
|Foyer| provides scripts to validate its output files by comparing against systems with known atom types (e.g., those determined by hand or reference molecules provided by a force field developer).
Output validation requires (1) system(s) with known, validated atom types and (2) the force field XML file.
The known systems are read into |Foyer| and atom types are determined using the rules in the XML file.
The atom types generated by |Foyer| are then compared against the known atom-typed system(s).
The |pytest|\cite{pytest} library is used to provide a clear, descriptive output of the results of these validation tests.
Implementing output validation tests is particularly useful to force field developers as they ensure that the desired output is retained if a force field file is evolved through the addition of new atom type definitions or merged with a separate force field file.
The utility of these validation checks relies not only upon providing accurate reference systems, but also a sufficient variety of test systems that encompass all defined atom types, as discussed previously.
An example of such a validation test suite is provided as part of the |Foyer| template repository \cite{foyer_template}.

\section{Usage Examples}
Here, a basic overview of the usage of |Foyer| is provided.
Readers are also directed to the GitHub project repository \cite{foyer} and tutorial repository \cite{foyer_tutorials} for more thorough and current usage examples.
Consider constructing a bulk system of ethane molecules and applying the OPLS force field.
Listing \ref{lst:ethane} shows a simple |mBuild|\cite{} script to load an ethane molecule and fill a 2nm x 2nm x 2nm box with 100 molecules.
This defines the system's chemical topology to which the force field will be applied.
As input, the force field file is identical to Listing \ref{lst:xml_format} but with the |<AtomType>| information from Listing \ref{lst:methyl_atomtypes}, as Listing \ref{lst:methyl_atomtypes} includes the usage rule definitions.
Listing \ref{lst:ethane} demonstrates two different syntaxes for applying a force field using |Foyer| and saving the output, in this case to the file format required by GROMACS.
The second option allows different forcefields to be applied to different topologies in the system.
Listing \ref{lst:combined_ff} shows an example of creating two separate chemical topologies in the system, and applying two different force field files to each.
The two atom-typed structures that result (|ethane_fluid| and |silica_substrate|) are then combined using a simple $+$ operator and saved to any format supported by ParmEd.
Note that if the surface and polymers were bonded together (e.g., to create a surface-bound monolayer), the force field files would need to be combined into a single XML document.

\begin{minipage}{\linewidth}
\begin{lstlisting}[caption={Script to fill a box with ethane and apply the OPLS-AA force field to the system.}, label=lst:ethane, language=python, basicstyle=\footnotesize\ttfamily]
import mbuild as mb
from mbuild.examples import Ethane
from foyer.test.utils import get_fn
from foyer import Forcefield

### Approach 1 ###
# Create the chemical topology
ethane_fluid = mb.fill_box(compound=Ethane(), n_compounds=100, box=[2, 2, 2])
# Apply and save the topology
ethane_fluid.save('ethane-box.top', forcefield_files=get_fn('oplsaa_alkane.xml'))
ethane_fluid.save('ethane-box.gro')

### Approach 2 ###
# Create the chemical topology
ethane_fluid = mb.fill_box(compound=Ethane(), n_compounds=100, box=[2, 2, 2])
# Load the forcefield
opls_alkane = Forcefield(forcefield_files=get_fn('oplsaa_alkane.xml'))
# Apply the forcefield to atom-type
ethane_fluid = opls_alkane.apply(ethane_fluid)
# Save the atom-typedsystem
ethane_fluid.save('ethane-box.top', overwrite=True)
ethane_fluid.save('ethane-box.gro', overwrite=True)

\end{lstlisting}
\end{minipage}

\begin{minipage}{\linewidth}
\begin{lstlisting}[caption={Script to build a system with an amorphous silica substrate in contact with a bulk ethane system and apply a different force field to the substrate and fluid respectively.}, label=lst:combined_ff, language=Python]
from foyer import Forcefield
from foyer.test.utils import get_fn
import mbuild as mb
from mbuild.examples import Ethane
from mbuild.lib.atoms import H
from mbuild.lib.bulk_materials import AmorphousSilica

# Create a silica substrate, capping surface oxygens with hydrogen
silica = mb.SilicaInterface(bulk_silica=AmorphousSilica())
silica_substrate = mb.Monolayer(surface=silica, chains=H(), guest_port_name='up')
# Determine the box dimensions dictated by the silica substrate
box = mb.Box(mins=[0, 0, max(silica.xyz[:,2])], maxs=silica.periodicity + [0, 0, 4])
# Fill the box with ethane
ethane_fluid = mb.fill_box(compound=Ethane(), n_compounds=200, box=box)
# Load the forcefields
opls_silica = Forcefield(forcefield_files=get_fn('opls-silica.xml'))
opls_alkane = Forcefield(forcefield_files=get_fn('oplsaa_alkane.xml'))
# Apply the forcefields
silica_substrate = opls_silica.apply(silica_substrate)
ethane_fluid = opls_alkane.apply(ethane_fluid)
# Merge the two topologies
system = silica_substrate + ethane_fluid
# Save the atom-typed system
system.save('ethane-silica.top')
system.save('ethane-silica.gro')
\end{lstlisting}
\end{minipage}

\section{Promoting Reproducible Force Field Dissemination}
Foyer force field files and associated documentation, examples, and validation tests, can be readily developed and distributed using standard software development approaches to improve quality and reproducibility.
For example, the common git + GitHub/Bitbucket based distribution process allows force field creators to disseminate their force field files and associated content to the public via a version controlled repository that can be referenced from relevant publications.
In this approach, a specific version of the force field used in a publication can be tagged in the git repository and a reference to this tagged version provided in the manuscript, allowing for a clear reference to the exact parameters and usage rules employed in the work.
Other services, such as Zenodo \cite{zenodo}, can additionally provide a digital object identifier (DOI) for the tagged record and a snapshot of the content of the archive.
A variety of other features of this standard software development process translate well to force field development.
Version control systems like git are designed to facilitate distributed, collaborative software development and allow for changes to the files in the repository to be easily tracked in a transparent manner.
For example, as a force field is evolved or corrected, revisions can be easily tracked, including the author(s) responsible for the changes, and the specific differences between force field versions clearly identified using standard tools such as DIFF and through the use of descriptive ``commit'' statements as the content of the repository is changed.
Whenever the developers wish to, they can create a new release of the force field that, as noted above, can be tagged or provided with a citable DOI.
Verification and validation of a force field can also be simplified by using this software design approach, i.e., by implementing automated testing tools that can perform checks on every new iteration (i.e., commit) of the force field content to ensure errors are not introduced as the force field is changed.

To promote these practices, we have created a template git repository on GitHub which contains the basic framework needed to create, test, and publish a new force field as well as a guided tutorial that introduces users to the SMARTS based atom-typing scheme \cite{foyer_template}.
This process was successfully used in recent work that derived force field parameters for perfluoropolyethers\cite{Black2017}, a novel lubricant class.
The force field was published in conjunction with the manuscript and made available on GitHub (\url{https://github.com/mosdef-hub/forcefield_perfluoroethers}).
The specific version of the force field at time of publication is citable via a separate DOI\cite{PFPEs}.
Any adjustments or improvements to the force field can be released under a new DOI, while the old one would still exist and point to the originally published force field in order to maintain provenance.

\subsection{Atom type DOI labels}
While automated atom-typing and the containment of atomtypes and force field parameters within a single file helps reduce user error and promotes reproducibility, users also require knowledge of the original source of parameters, in order to ensure proper citation and validation that the parameters are appropriate for their system of interest.
|Foyer| achieves this goal by adding a |doi| attribute  to each |Type| definition within the |AtomTypes| block of a force field XML.
Listing \ref{lst:doi} shows the same atom-type definition for the OPLS-AA methyl carbon as in Listing \ref{lst:methyl_atomtypes} with the additional |doi| attribute providing the DOI to the original source where parameters for this atom type were derived.

\begin{minipage}{\linewidth}
\begin{lstlisting}[caption={Atom type definition for a methyl carbon tagged with the source DOI}, label=lst:doi, language=xml, frame=tlrb]
<ForceField>
 <AtomTypes>
  <Type name="opls_135" class="CT" element="C" mass="12.01100"\\
    def="[C;X4](C)(H)(H)H" desc="alkane CH3" \\
    doi="10.1021/ja9621760"/>
 </AtomTypes>
</ForceField>
\end{lstlisting}
\end{minipage}

This feature eliminates ambiguity concerning the origin of parameters for a particular atom type.
Furthermore, |Foyer| automatically logs associations between DOIs and atomtypes during the atom-typing process, providing a BibTeX file featuring the full citation for the sources of all parameters applied to a particular system, along with additional notes detailing precisely which atomtypes are contained within each source.
For example, Listing \ref{lst:bibtex} shows the BibTeX file generated for a nitropropane molecule using the OPLS-AA force field, which includes all reference information as well as notes describing which atom type parameters were obtained from each reference.

\begin{minipage}{\linewidth}
\begin{lstlisting}[caption={BibTeX file generated during atom-typing of nitropropane using the OPLS-AA force field (modified with line breaks for readability).}, label=lst:bibtex, basicstyle=\footnotesize\ttfamily, frame=tlrb]
@article{Price_2001,
    doi = {10.1002/jcc.1092},
    url = {https://doi.org/10.1002%2Fjcc.1092},
    year = 2001,
    publisher = {Wiley-Blackwell},
    volume = {22},
    number = {13},
    pages = {1340--1352},
    author = {Melissa L. P. Price and Dennis Ostrovsky and William L. Jorgensen},
    title = {Gas-phase and liquid-state properties of esters, nitriles, and nitro
        compounds with the {OPLS}-{AA} force field},
    journal = {Journal of Computational Chemistry},
    note = {Parameters for atom types: opls_761, opls_760, opls_764, opls_763}
}
@article{Jorgensen_1996,
    doi = {10.1021/ja9621760},
    url = {https://doi.org/10.1021%2Fja9621760},
    year = 1996,
    month = {jan},
    publisher = {American Chemical Society ({ACS})},
    volume = {118},
    number = {45},
    pages = {11225--11236},
    author = {William L. Jorgensen and David S. Maxwell and Julian Tirado-Rives},
    title = {Development and Testing of the {OPLS} All-Atom Force Field
        on Conformational Energetics and Properties of Organic Liquids},
    journal = {Journal of the American Chemical Society},
    note = {Parameters for atom types: opls_135, opls_140, opls_136}
}
\end{lstlisting}
\end{minipage}

\section{Conclusion}
|Foyer| is a Python package for atom-typing and parameterization of molecular models, yielding an output directly usable by common molecular simulation engines.
|Foyer| extends the XML force field format introduced by |OpenMM| with an additional field for SMARTS definitions that define the chemical context of atom types in a single file that is both human and machine readable.
|Foyer| uses these SMARTS definitions to assign the appropriate atom types and provides an |overrides| syntax for defining rule precedence that is independent of the order of the rule definitions within the file.
|Foyer| force fields can be used along with standard software development practices for version control and DOI tagging to provide a complete record of force field development and links that researchers can include within a publication leading to the exact force field parameters used for a particular study. 
The extension of the XML file format to additional include DOI tags for the source of each parameter, further increases the clarity of force field usage. 
By promoting best practices in terms of encoding parameter usage and disseminating force fields, we believe that |Foyer| can play a critical role in improving reproducibility in the molecular simulation community.

\section{Acknowledgments}
This material is based upon work supported by the National Science Foundation under Grant No. OCI-1047828.
We also acknowledge the National Energy Research Supercomputing Center, which is supported by the Office of Science of the U.S. Department of Energy under Contract No.  DE-AC02-05CH11231.

\bibliography{foyer.bib}

\providecommand*\mcitethebibliography{\thebibliography}
\csname @ifundefined\endcsname{endmcitethebibliography}
  {\let\endmcitethebibliography\endthebibliography}{}
\begin{mcitethebibliography}{45}
\providecommand*\natexlab[1]{#1}
\providecommand*\mciteSetBstSublistMode[1]{}
\providecommand*\mciteSetBstMaxWidthForm[2]{}
\providecommand*\mciteBstWouldAddEndPuncttrue
  {\def\EndOfBibitem{\unskip.}}
\providecommand*\mciteBstWouldAddEndPunctfalse
  {\let\EndOfBibitem\relax}
\providecommand*\mciteSetBstMidEndSepPunct[3]{}
\providecommand*\mciteSetBstSublistLabelBeginEnd[3]{}
\providecommand*\EndOfBibitem{}
\mciteSetBstSublistMode{f}
\mciteSetBstMaxWidthForm{subitem}{(\alph{mcitesubitemcount})}
\mciteSetBstSublistLabelBeginEnd
  {\mcitemaxwidthsubitemform\space}
  {\relax}
  {\relax}

\bibitem[Weiner et~al.(1984)Weiner, Kollman, Case, Singh, Ghio, Alagona,
  Profeta, and Weinerl]{Weiner1984}
Weiner,~S.~J.; Kollman,~P.~A.; Case,~D.~A.; Singh,~U.~C.; Ghio,~C.;
  Alagona,~G.; Profeta,~S.; Weinerl,~P. \emph{Journal of the American Chemical
  Society} \textbf{1984}, \emph{106}, 765--784\relax
\mciteBstWouldAddEndPuncttrue
\mciteSetBstMidEndSepPunct{\mcitedefaultmidpunct}
{\mcitedefaultendpunct}{\mcitedefaultseppunct}\relax
\EndOfBibitem
\bibitem[MacKerell et~al.(2000)MacKerell, Banavali, and Foloppe]{MacKerell2000}
MacKerell,~A.~D.; Banavali,~N.; Foloppe,~N. \emph{Biopolymers} \textbf{2000},
  \emph{56}, 257--65\relax
\mciteBstWouldAddEndPuncttrue
\mciteSetBstMidEndSepPunct{\mcitedefaultmidpunct}
{\mcitedefaultendpunct}{\mcitedefaultseppunct}\relax
\EndOfBibitem
\bibitem[Jorgensen et~al.(1996)Jorgensen, Maxwell, and
  Tirado-Rives]{Jorgensen1996}
Jorgensen,~W.~L.; Maxwell,~D.~S.; Tirado-Rives,~J. \emph{Journal of the
  American Chemical Society} \textbf{1996}, \emph{118}, 11225--11236\relax
\mciteBstWouldAddEndPuncttrue
\mciteSetBstMidEndSepPunct{\mcitedefaultmidpunct}
{\mcitedefaultendpunct}{\mcitedefaultseppunct}\relax
\EndOfBibitem
\bibitem[Siepmann et~al.(1993)Siepmann, Karaborni, and
  Smit]{siepmann1993simulating}
Siepmann,~J.~I.; Karaborni,~S.; Smit,~B. \emph{Nature} \textbf{1993},
  \emph{365}, 330\relax
\mciteBstWouldAddEndPuncttrue
\mciteSetBstMidEndSepPunct{\mcitedefaultmidpunct}
{\mcitedefaultendpunct}{\mcitedefaultseppunct}\relax
\EndOfBibitem
\bibitem[Potoff and Siepmann(2001)Potoff, and Siepmann]{Potoff2001}
Potoff,~J.~J.; Siepmann,~J.~I. \emph{AIChE Journal} \textbf{2001}, \emph{47},
  1676--1682\relax
\mciteBstWouldAddEndPuncttrue
\mciteSetBstMidEndSepPunct{\mcitedefaultmidpunct}
{\mcitedefaultendpunct}{\mcitedefaultseppunct}\relax
\EndOfBibitem
\bibitem[Sun(1998)]{Sun1998}
Sun,~H. \emph{Journal of Physical Chemistry} \textbf{1998}, \emph{5647},
  7338--7364\relax
\mciteBstWouldAddEndPuncttrue
\mciteSetBstMidEndSepPunct{\mcitedefaultmidpunct}
{\mcitedefaultendpunct}{\mcitedefaultseppunct}\relax
\EndOfBibitem
\bibitem[Oostenbrink et~al.(2004)Oostenbrink, Villa, Mark, and {Van
  Gunsteren}]{Oostenbrink2004}
Oostenbrink,~C.; Villa,~A.; Mark,~A.~E.; {Van Gunsteren},~W.~F. \emph{Journal
  of Computational Chemistry} \textbf{2004}, \emph{25}, 1656--1676\relax
\mciteBstWouldAddEndPuncttrue
\mciteSetBstMidEndSepPunct{\mcitedefaultmidpunct}
{\mcitedefaultendpunct}{\mcitedefaultseppunct}\relax
\EndOfBibitem
\bibitem[Gro()]{GromacsOPLS}
{Gromacs OPLS Atom Types}.
  \url{https://github.com/gromacs/gromacs/blob/e131e1d16c589fded5cad47bbd52b010d59c80a7/share/top/oplsaa.ff/atomtypes.atp}\relax
\mciteBstWouldAddEndPuncttrue
\mciteSetBstMidEndSepPunct{\mcitedefaultmidpunct}
{\mcitedefaultendpunct}{\mcitedefaultseppunct}\relax
\EndOfBibitem
\bibitem[Sandve et~al.(2013)Sandve, Nekrutenko, Taylor, and
  Hovig]{sandve2013ten}
Sandve,~G.~K.; Nekrutenko,~A.; Taylor,~J.; Hovig,~E. \emph{PLoS computational
  biology} \textbf{2013}, \emph{9}, e1003285\relax
\mciteBstWouldAddEndPuncttrue
\mciteSetBstMidEndSepPunct{\mcitedefaultmidpunct}
{\mcitedefaultendpunct}{\mcitedefaultseppunct}\relax
\EndOfBibitem
\bibitem[Bush and Sheridan(1993)Bush, and Sheridan]{Bush1993}
Bush,~B.~L.; Sheridan,~R.~P. \emph{Journal of Chemical Information and Computer
  Sciences} \textbf{1993}, \emph{33}, 756--762\relax
\mciteBstWouldAddEndPuncttrue
\mciteSetBstMidEndSepPunct{\mcitedefaultmidpunct}
{\mcitedefaultendpunct}{\mcitedefaultseppunct}\relax
\EndOfBibitem
\bibitem[Sch{\"{u}}ttelkopf and {Van Aalten}(2004)Sch{\"{u}}ttelkopf, and {Van
  Aalten}]{Schuttelkopf2004a}
Sch{\"{u}}ttelkopf,~A.~W.; {Van Aalten},~D. M.~F. \emph{Acta Crystallographica
  Section D: Biological Crystallography} \textbf{2004}, \emph{60},
  1355--1363\relax
\mciteBstWouldAddEndPuncttrue
\mciteSetBstMidEndSepPunct{\mcitedefaultmidpunct}
{\mcitedefaultendpunct}{\mcitedefaultseppunct}\relax
\EndOfBibitem
\bibitem[Wang et~al.(2006)Wang, Wang, Kollman, and Case]{Wang2006}
Wang,~J.; Wang,~W.; Kollman,~P.~a.; Case,~D.~a. \emph{Journal of Molecular
  Graphics {\&} Modelling} \textbf{2006}, \emph{25}, 247--60\relax
\mciteBstWouldAddEndPuncttrue
\mciteSetBstMidEndSepPunct{\mcitedefaultmidpunct}
{\mcitedefaultendpunct}{\mcitedefaultseppunct}\relax
\EndOfBibitem
\bibitem[Ribeiro et~al.(2008)Ribeiro, Horta, and {De Alencastro}]{Ribeiro2008}
Ribeiro,~A. A. S.~T.; Horta,~B. A.~C.; {De Alencastro},~R.~B. \emph{Journal of
  the Brazilian Chemical Society} \textbf{2008}, \emph{19}, 1433--1435\relax
\mciteBstWouldAddEndPuncttrue
\mciteSetBstMidEndSepPunct{\mcitedefaultmidpunct}
{\mcitedefaultendpunct}{\mcitedefaultseppunct}\relax
\EndOfBibitem
\bibitem[Malde et~al.(2011)Malde, Zuo, Breeze, Stroet, Poger, Nair,
  Oostenbrink, and Mark]{Malde2011}
Malde,~A.~K.; Zuo,~L.; Breeze,~M.; Stroet,~M.; Poger,~D.; Nair,~P.~C.;
  Oostenbrink,~C.; Mark,~A.~E. \emph{Journal of Chemical Theory and
  Computation} \textbf{2011}, \emph{7}, 4026--4037\relax
\mciteBstWouldAddEndPuncttrue
\mciteSetBstMidEndSepPunct{\mcitedefaultmidpunct}
{\mcitedefaultendpunct}{\mcitedefaultseppunct}\relax
\EndOfBibitem
\bibitem[Vanommeslaeghe and MacKerell(2012)Vanommeslaeghe, and
  MacKerell]{Vanommeslaeghe2012}
Vanommeslaeghe,~K.; MacKerell,~a.~D. \emph{Journal of Chemical Information and
  Modeling} \textbf{2012}, \emph{52}, 3144--54\relax
\mciteBstWouldAddEndPuncttrue
\mciteSetBstMidEndSepPunct{\mcitedefaultmidpunct}
{\mcitedefaultendpunct}{\mcitedefaultseppunct}\relax
\EndOfBibitem
\bibitem[Yesselman et~al.(2012)Yesselman, Price, Knight, and
  Brooks]{Yesselman2012}
Yesselman,~J.~D.; Price,~D.~J.; Knight,~J.~L.; Brooks,~C.~L. \emph{Journal of
  Computational Chemistry} \textbf{2012}, \emph{33}, 189--202\relax
\mciteBstWouldAddEndPuncttrue
\mciteSetBstMidEndSepPunct{\mcitedefaultmidpunct}
{\mcitedefaultendpunct}{\mcitedefaultseppunct}\relax
\EndOfBibitem
\bibitem[Weininger(1988)]{weiniger1988smiles}
Weininger,~D. \emph{Journal of Chemical Information and Computer Sciences}
  \textbf{1988}, \emph{28}, 31--36\relax
\mciteBstWouldAddEndPuncttrue
\mciteSetBstMidEndSepPunct{\mcitedefaultmidpunct}
{\mcitedefaultendpunct}{\mcitedefaultseppunct}\relax
\EndOfBibitem
\bibitem[in't Veld()]{emc}
in't Veld,~P.~J. {EMC: Enhanced Monte Carlo; A multi-purpose modular and easily
  extendable solution to molecular and mesoscale simulations}.
  \url{http://montecarlo.sourceforge.net}\relax
\mciteBstWouldAddEndPuncttrue
\mciteSetBstMidEndSepPunct{\mcitedefaultmidpunct}
{\mcitedefaultendpunct}{\mcitedefaultseppunct}\relax
\EndOfBibitem
\bibitem[Mobley et~al.(2018)Mobley, Bannan, Rizzi, Bayly, Chodera, Lim, Lim,
  Beauchamp, Shirts, Gilson, and Eastman]{Mobley2018}
Mobley,~D.; Bannan,~C.~C.; Rizzi,~A.; Bayly,~C.~I.; Chodera,~J.~D.; Lim,~V.~T.;
  Lim,~N.~M.; Beauchamp,~K.~A.; Shirts,~M.~R.; Gilson,~M.~K.; Eastman,~P.~K.
  \emph{bioRxiv} \textbf{2018}, \relax
\mciteBstWouldAddEndPunctfalse
\mciteSetBstMidEndSepPunct{\mcitedefaultmidpunct}
{}{\mcitedefaultseppunct}\relax
\EndOfBibitem
\bibitem[Day()]{DaylightSMARTS}
{Daylight Theory: SMARTS - A Language for Describing Molecular Patterns}.
  \url{http://www.daylight.com/dayhtml/doc/theory/theory.smarts.html}\relax
\mciteBstWouldAddEndPuncttrue
\mciteSetBstMidEndSepPunct{\mcitedefaultmidpunct}
{\mcitedefaultendpunct}{\mcitedefaultseppunct}\relax
\EndOfBibitem
\bibitem[foy()]{foyer}
{Foyer GitHub Repository}. \url{http://github.com/mosdef-hub/foyer}\relax
\mciteBstWouldAddEndPuncttrue
\mciteSetBstMidEndSepPunct{\mcitedefaultmidpunct}
{\mcitedefaultendpunct}{\mcitedefaultseppunct}\relax
\EndOfBibitem
\bibitem[par()]{parmed}
{ParmEd}. \url{http://parmed.github.io/ParmEd/html/index.html}\relax
\mciteBstWouldAddEndPuncttrue
\mciteSetBstMidEndSepPunct{\mcitedefaultmidpunct}
{\mcitedefaultendpunct}{\mcitedefaultseppunct}\relax
\EndOfBibitem
\bibitem[Eastman et~al.(2013)Eastman, Friedrichs, Chodera, Radmer, Bruns, Ku,
  Beauchamp, Lane, Wang, Shukla, Tye, Houston, Stich, Klein, Shirts, and
  Pande]{Eastman2013}
Eastman,~P. et~al.  \emph{Journal of Chemical Theory and Computation}
  \textbf{2013}, \emph{9}, 461--469\relax
\mciteBstWouldAddEndPuncttrue
\mciteSetBstMidEndSepPunct{\mcitedefaultmidpunct}
{\mcitedefaultendpunct}{\mcitedefaultseppunct}\relax
\EndOfBibitem
\bibitem[Sallai et~al.(2014)Sallai, Varga, Toth, Iacovella, Klein, McCabe,
  Ledeczi, and Cummings]{Sallai2014}
Sallai,~J.; Varga,~G.; Toth,~S.; Iacovella,~C.; Klein,~C.; McCabe,~C.;
  Ledeczi,~A.; Cummings,~P.~T. \emph{Procedia Computer Science} \textbf{2014},
  \emph{29}, 2034--2044\relax
\mciteBstWouldAddEndPuncttrue
\mciteSetBstMidEndSepPunct{\mcitedefaultmidpunct}
{\mcitedefaultendpunct}{\mcitedefaultseppunct}\relax
\EndOfBibitem
\bibitem[Klein et~al.(2016)Klein, Sallai, Jones, Iacovella, McCabe, and
  Cummings]{Klein2016}
Klein,~C.; Sallai,~J.; Jones,~T.~J.; Iacovella,~C.~R.; McCabe,~C.;
  Cummings,~P.~T. In \emph{Foundations of Molecular Modeling and Simulation.
  Molecular Modeling and Simulation (Applications and Perspectives)};
  Snurr,~R.~Q., Adjiman,~C.~S., Kofke,~D.~A., Eds.; Springer, Singapore:
  Singapore, 2016; pp 79--92\relax
\mciteBstWouldAddEndPuncttrue
\mciteSetBstMidEndSepPunct{\mcitedefaultmidpunct}
{\mcitedefaultendpunct}{\mcitedefaultseppunct}\relax
\EndOfBibitem
\bibitem[mBu()]{mBuild}
{mBuild, GitHub Repository}. \url{https://github.com/mosdef-hub/mbuild}\relax
\mciteBstWouldAddEndPuncttrue
\mciteSetBstMidEndSepPunct{\mcitedefaultmidpunct}
{\mcitedefaultendpunct}{\mcitedefaultseppunct}\relax
\EndOfBibitem
\bibitem[sma()]{smarts}
{SMARTS - A Language for Describing Molecular Patterns}.
  \url{http://www.daylight.com/dayhtml/doc/theory/theory.smarts.html}\relax
\mciteBstWouldAddEndPuncttrue
\mciteSetBstMidEndSepPunct{\mcitedefaultmidpunct}
{\mcitedefaultendpunct}{\mcitedefaultseppunct}\relax
\EndOfBibitem
\bibitem[Weininger(1988)]{weininger1988smiles}
Weininger,~D. \emph{Journal of Chemical Information and Computer Sciences}
  \textbf{1988}, \emph{28}, 31--36\relax
\mciteBstWouldAddEndPuncttrue
\mciteSetBstMidEndSepPunct{\mcitedefaultmidpunct}
{\mcitedefaultendpunct}{\mcitedefaultseppunct}\relax
\EndOfBibitem
\bibitem[Morgado et~al.(2013)Morgado, Black, Lewis, Iacovella, McCabe, Martins,
  and Filipe]{Morgado2013}
Morgado,~P.; Black,~J.; Lewis,~J.~B.; Iacovella,~C.~R.; McCabe,~C.; Martins,~L.
  F.~G.; Filipe,~E. J.~M. \emph{Fluid Phase Equilibria} \textbf{2013},
  \emph{358}, 161--165\relax
\mciteBstWouldAddEndPuncttrue
\mciteSetBstMidEndSepPunct{\mcitedefaultmidpunct}
{\mcitedefaultendpunct}{\mcitedefaultseppunct}\relax
\EndOfBibitem
\bibitem[Oliphant(2015)]{Oliphant:2015:GN:2886196}
Oliphant,~T.~E. \emph{Guide to NumPy}, 2nd ed.; CreateSpace Independent
  Publishing Platform: USA, 2015\relax
\mciteBstWouldAddEndPuncttrue
\mciteSetBstMidEndSepPunct{\mcitedefaultmidpunct}
{\mcitedefaultendpunct}{\mcitedefaultseppunct}\relax
\EndOfBibitem
\bibitem[Jones et~al.(2001--)Jones, Oliphant, Peterson, et~al. others]{scipy}
others,, et~al.  {SciPy}: Open source scientific tools for {Python}. 2001--;
  \url{http://www.scipy.org/}, [Online; accessed <today>]\relax
\mciteBstWouldAddEndPuncttrue
\mciteSetBstMidEndSepPunct{\mcitedefaultmidpunct}
{\mcitedefaultendpunct}{\mcitedefaultseppunct}\relax
\EndOfBibitem
\bibitem[Schult and Swart(2008)Schult, and Swart]{networkx}
Schult,~D.~A.; Swart,~P. Exploring network structure, dynamics, and function
  using NetworkX. 2008\relax
\mciteBstWouldAddEndPuncttrue
\mciteSetBstMidEndSepPunct{\mcitedefaultmidpunct}
{\mcitedefaultendpunct}{\mcitedefaultseppunct}\relax
\EndOfBibitem
\bibitem[Shirts et~al.(2017)Shirts, Klein, Swails, Yin, Gilson, Mobley, Case,
  and Zhong]{shirts2017lessons}
Shirts,~M.~R.; Klein,~C.; Swails,~J.~M.; Yin,~J.; Gilson,~M.~K.; Mobley,~D.~L.;
  Case,~D.~A.; Zhong,~E.~D. \emph{Journal of Computer-Aided Molecular Design}
  \textbf{2017}, \emph{31}, 147--161\relax
\mciteBstWouldAddEndPuncttrue
\mciteSetBstMidEndSepPunct{\mcitedefaultmidpunct}
{\mcitedefaultendpunct}{\mcitedefaultseppunct}\relax
\EndOfBibitem
\bibitem[Klein et~al.(2014)Klein, Sallai, Iacovella, McCabe, and
  Cummings]{Klein2014}
Klein,~C.; Sallai,~J.; Iacovella,~C.~R.; McCabe,~C.; Cummings,~P.~T. {Mbuild: A
  Hierarchical, Component Based Molecule Builder}. 2014\relax
\mciteBstWouldAddEndPuncttrue
\mciteSetBstMidEndSepPunct{\mcitedefaultmidpunct}
{\mcitedefaultendpunct}{\mcitedefaultseppunct}\relax
\EndOfBibitem
\bibitem[Cordella et~al.(2004)Cordella, Foggia, Sansone, and Vento]{vf2}
Cordella,~L.~P.; Foggia,~P.; Sansone,~C.; Vento,~M. \emph{IEEE transactions on
  pattern analysis and machine intelligence} \textbf{2004}, \emph{26},
  1367--1372\relax
\mciteBstWouldAddEndPuncttrue
\mciteSetBstMidEndSepPunct{\mcitedefaultmidpunct}
{\mcitedefaultendpunct}{\mcitedefaultseppunct}\relax
\EndOfBibitem
\bibitem[Wang et~al.(2004)Wang, Wolf, Caldwell, Kollman, and Case]{Wang2004}
Wang,~J.; Wolf,~R.~M.; Caldwell,~J.~W.; Kollman,~P.~A.; Case,~D.~A.
  \emph{Journal of Computational Chemistry} \textbf{2004}, \emph{25},
  1157--74\relax
\mciteBstWouldAddEndPuncttrue
\mciteSetBstMidEndSepPunct{\mcitedefaultmidpunct}
{\mcitedefaultendpunct}{\mcitedefaultseppunct}\relax
\EndOfBibitem
\bibitem[Martin and Siepmann(1998)Martin, and Siepmann]{Martin1998}
Martin,~M.~G.; Siepmann,~J.~I. \textbf{1998}, \emph{5647}, 2569--2577\relax
\mciteBstWouldAddEndPuncttrue
\mciteSetBstMidEndSepPunct{\mcitedefaultmidpunct}
{\mcitedefaultendpunct}{\mcitedefaultseppunct}\relax
\EndOfBibitem
\bibitem[tra()]{trappe_website}
{TraPPE Force Fields}.
  \url{http://www.chem.umn.edu/groups/siepmann/trappe/}\relax
\mciteBstWouldAddEndPuncttrue
\mciteSetBstMidEndSepPunct{\mcitedefaultmidpunct}
{\mcitedefaultendpunct}{\mcitedefaultseppunct}\relax
\EndOfBibitem
\bibitem[Krekel et~al.(2004)Krekel, Oliveira, Pfannschmidt, Bruynooghe,
  Laugher, and Bruhin]{pytest}
Krekel,~H.; Oliveira,~B.; Pfannschmidt,~R.; Bruynooghe,~F.; Laugher,~B.;
  Bruhin,~F. pytest. 2004; \url{https://github.com/pytest-dev/pytest}\relax
\mciteBstWouldAddEndPuncttrue
\mciteSetBstMidEndSepPunct{\mcitedefaultmidpunct}
{\mcitedefaultendpunct}{\mcitedefaultseppunct}\relax
\EndOfBibitem
\bibitem[foy()]{foyer_template}
{Foyer Force Field Template GitHub Repository}.
  \url{http://github.com/mosdef-hub/forcefield_template}\relax
\mciteBstWouldAddEndPuncttrue
\mciteSetBstMidEndSepPunct{\mcitedefaultmidpunct}
{\mcitedefaultendpunct}{\mcitedefaultseppunct}\relax
\EndOfBibitem
\bibitem[foy()]{foyer_tutorials}
{Foyer Tutorial GitHub Repository}.
  \url{http://github.com/mosdef-hub/foyer_tutorials}\relax
\mciteBstWouldAddEndPuncttrue
\mciteSetBstMidEndSepPunct{\mcitedefaultmidpunct}
{\mcitedefaultendpunct}{\mcitedefaultseppunct}\relax
\EndOfBibitem
\bibitem[zen()]{zenodo}
{Zenodo}. \url{https://zenodo.org/}\relax
\mciteBstWouldAddEndPuncttrue
\mciteSetBstMidEndSepPunct{\mcitedefaultmidpunct}
{\mcitedefaultendpunct}{\mcitedefaultseppunct}\relax
\EndOfBibitem
\bibitem[Black et~al.(2017)Black, Silva, Klein, Iacovella, Morgado, Martins,
  Filipe, and McCabe]{Black2017}
Black,~J.~E.; Silva,~G.~M.; Klein,~C.; Iacovella,~C.~R.; Morgado,~P.;
  Martins,~L.~F.; Filipe,~E.~J.; McCabe,~C. \emph{The Journal of Physical
  Chemistry B} \textbf{2017}, \relax
\mciteBstWouldAddEndPunctfalse
\mciteSetBstMidEndSepPunct{\mcitedefaultmidpunct}
{}{\mcitedefaultseppunct}\relax
\EndOfBibitem
\bibitem[Black et~al.(2017)Black, Silva, Klein, Iacovella, Morgado, Martins,
  Filipe, and McCabe]{PFPEs}
Black,~J.; Silva,~G.; Klein,~C.; Iacovella,~C.; Morgado,~P.; Martins,~L.;
  Filipe,~E.; McCabe,~C. OPLS-AA compatible parameters for perfluoroethers.
  2017; \url{https://doi.org/10.5281/zenodo.583310}\relax
\mciteBstWouldAddEndPuncttrue
\mciteSetBstMidEndSepPunct{\mcitedefaultmidpunct}
{\mcitedefaultendpunct}{\mcitedefaultseppunct}\relax
\EndOfBibitem
\end{mcitethebibliography}

\end{document}